\documentclass[useAMS,usenatbib]{mn2e}
\usepackage{txfonts}
\usepackage{graphicx}
\usepackage{textcomp}
\usepackage{natbib}
\usepackage{deluxetable}

\newcommand{\um}{\textmu m }
\newcommand{\uu}{\textmu m}

\newcommand{\uuJy}{\textmu Jy}

\newcommand{\oii}{[O {\sc ii}] }
\def\zspec{$z_{\rm{spec}}$ }

\def\zphot{$z_{\rm{phot}}$ }

\def\Dn{D$_n$(4000) }
\def\Dnu{D$_n$(4000)}
\def\logDn{$\log$D$_n$(4000) }
\def\logDnu{$\log$D$_n$(4000)}

\bibpunct[]{(}{)}{;}{a}{}{,}

\title[Stellar populations of $z$$\sim$0.9 galaxies]{SHARDS: stellar populations and star formation histories of a mass-selected sample of 0.65$<$$z$$<$1.1 galaxies}

\author[A. Hern\'an-Caballero et al.]{
Antonio Hern\'an-Caballero,$^1$
Almudena Alonso-Herrero,$^{1,2}$
Pablo G. P\'erez-Gonz\'alez,$^{3,4}$\newauthor
Nicol\'as Cardiel,$^3$
Antonio Cava,$^3$
Ignacio Ferreras,$^5$
Guillermo Barro,$^{3,6}$
Laurence Tresse,$^{7}$\newauthor
Emanuele Daddi,$^{8}$
Javier Cenarro,$^{9}$
Christopher J. Conselice,$^{10}$
Rafael Guzm\'an,$^{11}$\newauthor
and Jes\'us Gallego$^{3}$\\
$^{1}$Instituto de F\'isica de Cantabria, CSIC-UC, Avenida de los Castros s/n, 39005, Santander, Spain. E-mail: ahernan@ifca.unican.es\\
$^{2}$Augusto G. Linares Senior Research Fellow\\
$^{3}$Departamento de Astrof\'isica y Ciencias de la Atm\'osfera, Facultad de CC. F\'isicas, Universidad Complutense de Madrid, E-28040 Madrid, Spain\\
$^{4}$Associate Astronomer at Steward Observatory, The University of Arizona, Tucson, USA\\
$^{5}$Mullard Space Science Laboratory, University College London, Holmbury St Mary, Dorking, Surrey RH5 6NT, UK\\
$^{6}$UCO/Lick Observatory, Department of Astronomy and Astrophysics, University of California, Santa Cruz, CA 95064, USA\\
$^{7}$Aix Marseille Universit\'e, CNRS, LAM (Laboratoire d’Astrophysique de Marseille), UMR 7326, F-13388 Marseille, France\\
$^{8}$CEA, Laboratoire AIM, Irfu/SAp, F-91191 Gif-sur-Yvette, France\\
$^{9}$Centro de Estudios de F\'isica del Cosmos de Arag\'on, Plaza San Juan 1, Planta 2, E-44001 Teruel, Spain\\
$^{10}$School of Physics \& Astronomy, University of Nottingham, Nottingham NG7 2RD, UK\\
$^{11}$Department of Astronomy, University of Florida, 211 Bryant Space Science Center, Gainesville, FL 32611, USA
}

\begin{document}
\date{Accepted ........ Received ........;}

\pagerange{\pageref{firstpage}--\pageref{lastpage}} \pubyear{2013}

\maketitle

\label{firstpage}

\begin{abstract}
We report on results from the analysis of a stellar mass-selected ($\log$(M$_*$/M$_\odot$) $\ge$ 9.0) sample of 1644 galaxies at 0.65$<$$z$$<$1.1 with ultra-deep (m$_{AB}$$<$26.5) optical medium-band (R$\sim$50) photometry from the Survey for High-z Absorption Red and Dead Sources (SHARDS).
The spectral resolution of SHARDS allows us to consistently measure the strength of the 4000 \AA{} spectral break [\Dnu, an excellent age indicator for the stellar populations of quiescent galaxies] for all galaxies at $z$$\sim$0.9 down to $\log$(M$_*$/M$_\odot$)$\sim$9. The \Dn index cannot be resolved from broad-band photometry, and measurements from optical spectroscopic surveys are typically limited to galaxies at least $\times10$ more massive. 
When combined with the rest-frame U-V colour, (U-V)$_r$, \Dn provides a powerful diagnostic of the extinction affecting the stellar population that is relatively insensitive to degeneracies with age, metallicity or star formation history.
We use this novel approach to estimate de-reddened colours and light-weighted stellar ages for individual sources. We explore the relationships linking stellar mass, (U-V)$_r$, and \Dn for the sources in the sample, and compare them to those found in local galaxies. The main results are: a) both \Dn and (U-V)$_r$ correlate with M$_*$. The dispersion in \Dn values at a given M$_*$ increases with M$_*$, while the dispersion for (U-V)$_r$ decreases due to the higher average extinction prevalent in massive star-forming galaxies. b) for massive galaxies, we find a smooth transition between the blue cloud and red sequence in the intrinsic U-V colour, in contrast with other recent results. c) at a fixed stellar age, we find a positive correlation between extinction and stellar mass. d) the fraction of sources with declining or halted star formation increases steeply with the stellar mass, from $\sim$5\% at $\log$(M$_*$/M$_\odot$)=9.0--9.5 to $\sim$80\% at $\log$(M$_*$/M$_\odot$)$>$11, in agreement with downsizing scenarios.
\end{abstract}


\section{Introduction} 

\begin{figure*}
\begin{center}
\includegraphics[width=17.5cm]{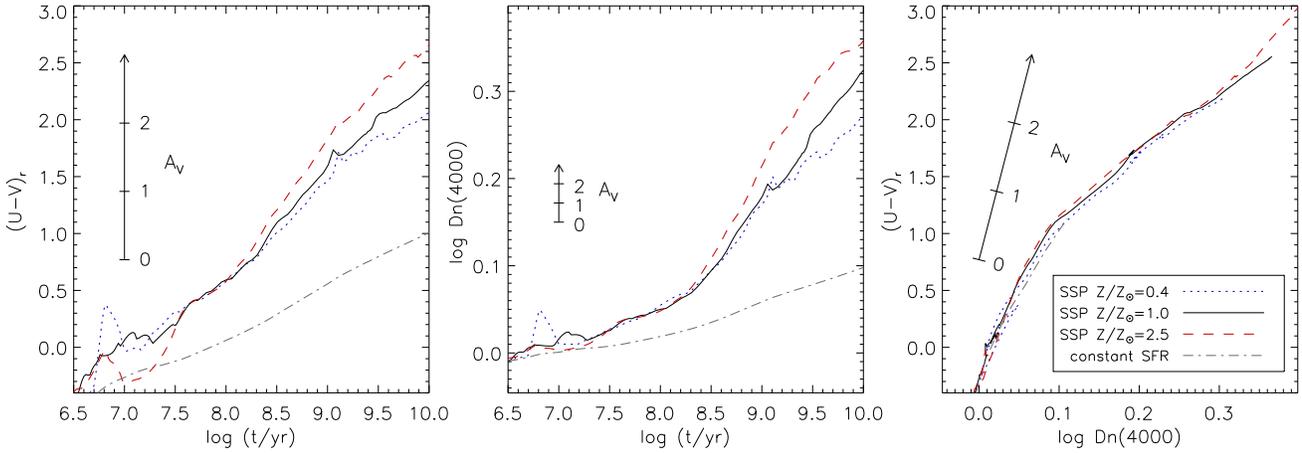}
\end{center}
\caption[]{Dependency of the rest-frame U-V colour, (U-V)$_r$, and the 4000 \AA{} index \citep[\Dnu;][]{Balogh99} with stellar age, extinction and metallicity. The tracks represent the evolution with time of the (U-V)$_r$ and $\log$\Dn colours of an instantaneous burst for A$_V$=0 and metallicity Z/Z$_\odot$ = 0.4, 1.0, and 2.5. The time evolution of a galaxy with Z/Z$_\odot$ = 1.0 and constant SFR is also shown for comparison. Model tracks are computed using GALAXEV \citep{Bruzual03} with the STELIB stellar library and Padova1994 isochrones \citep{Bertelli94} assuming a \citet{Salpeter55} initial mass function (IMF). Extinction produces a reddening that is proportional to A$_V$ for both (U-V)$_r$ and $\log$\Dn. The segmented arrows represent the effect of A$_V$ = 1, 2, and 3 magnitudes assuming a \citet{Draine03} extinction law with Milky Way grain size distribution \citep{Weingartner01} and R$_V$ = 3.1.\label{models}}
\end{figure*}

In the last decade it has been established that the colours and morphologies of galaxies show bimodal distributions at least up to $z$$\sim$3 \citep[e.g.][]{Blanton03,Bell04,Borch06,DeLucia07,Ilbert09,Williams09,Brammer09,Xue10,Whitaker11}, and that they correlate strongly with luminosity \citep[e.g.][]{Loveday92,Marzke94,Lin96,Christlein00} and stellar mass \citep{Xue10}.

In the local Universe, massive galaxies (M$_*$$>$10$^{11}$ M$_\odot$) are typically spheroids with old stellar populations, which define the so-called red sequence in a color-magnitude diagram (CMD), while the less massive ones (mostly discs) have young stellar populations and reside in a more spread blue cloud \citep{Strateva01,Kauffmann03b,Baldry04}. Galaxies with intermediate colours (located in the so-called green valley) are less numerous and are considered to be either dusty star-forming systems, or galaxies transitioning from the blue cloud to the red sequence after switching off their star formation.

At higher redshift, only increasingly massive galaxies belong to the red sequence, and the
total mass contained in it at $z$$\sim$1 is only half the local value, while the mass in the blue cloud has remained nearly constant \citep[e.g.][]{Arnouts07,Lotz08,Stutz08,Taylor09,Ilbert10,Dominguez11}.

To interpret these observations, a picture has emerged in which galaxy evolution is strongly coupled to the mass of the galactic halo and its baryon content. In the so-called downsizing scenario \citep{Cowie96}, the most massive galaxies are formed first and the star formation continues in lower mass systems until later epochs \citep{Heavens04,Bauer05,Perez-Gonzalez05,Perez-Gonzalez08,Bundy06,Tresse07}.

Quantifying the downsizing process requires accurate stellar mass estimates for large and representative samples of galaxies, as well as a reliable determination of the star formation history (SFH) of individual galaxies. 
For stellar masses, fitting UV to near-infrared SEDs with stellar population synthesis (SPS) models provides accurate results, with estimated uncertainties around 0.2 dex \citep[e.g.][]{Perez-Gonzalez08,Marchesini09,Barro11}.
Unfortunately, there is some degeneracy among stellar age, obscuration, and metallicity, which in turn introduces uncertainties in age estimates based on SED fitting. Although NIR photometry can break to some extent this degeneracy and there has been some success at obtaining de-reddened color indices \citep[e.g.][]{Brammer09,Cardamone10}, more specific stellar age indicators (like the strength of the 4000 \AA{} spectral break) are required. 

The 4000 \AA{} index, \Dnu, is an excellent age indicator for the stellar populations of quiescent galaxies \citep{Balogh99,Kauffmann03a,Kauffmann03b,Kriek06,Moresco10}. Since \Dn is much less affected by extinction compared to the U-V colour (see Figure \ref{models}),
it has also been used to separate star-forming and passively evolving galaxies \citep[e.g.][]{Vergani08}.

Large spectroscopic surveys, in particular the Sloan Digital Sky Survey, have provided a clear picture of the interdependency of stellar mass, surface density, and star formation history \citep{Kauffmann03b,Heavens04, Panter07, CidFernandes07}. Nevertheless, the spectra of stellar populations forming all their mass at $z$$\sim$1 or higher are indistinguishable in the local Universe, and to explore the entire SFH of local galaxies it is necessary to observe their higher redshift analogs. 
A typical M$_*$ = 10$^{10}$ M$_\odot$ galaxy at $z$=1 has m$\sim$24 (AB) in the I band, 
which matches the current faint limit for the deepest spectroscopic surveys, such as VVDS-deep \citep{LeFevre04,LeFevre05}.
Consequently, spectroscopic surveys favour the most massive galaxies, while the analysis of the stellar populations in lower-mass systems has remained constrained by the limitations of broadband SEDs. 

Recently, \citet{Kriek11} demonstrated that the 4000\nolinebreak{} \AA{} break can be successfully measured in average SEDs from medium-band photometry.
In this work we use the ultra-deep (m$_{AB}$$<$26.5, 3$\sigma$) optical medium-band spectro-photometry provided by the Survey for High-z Absorption Red and Dead Sources \citep[SHARDS;][]{Perez-Gonzalez13} in combination with accurate photometric and spectroscopic redshifts and stellar masses from the Rainbow\footnote{https://rainbowx.fis.ucm.es} database \citep[][; hereafter PG08]{Perez-Gonzalez08} to analyse the stellar populations and star formation histories of a mass-selected sample of galaxies at 0.65$<$$z$$<$1.1.

SHARDS is imaging an area of 141 arcmin$^2$ in the GOODS-North field with GTC/OSIRIS in 24 contiguous medium-band (R$\sim$50) filters covering the spectral range 500--950 nm. At the time of this writing, the entire survey area has been observed in the 15 filters covering the spectral range 636--883 nm, and one half of the area has also been observed in the F619W17 filter (central wavelength: 619 nm, FWHM: 17 nm). 

We measure the 4000 \AA{} index in the SHARDS SED of individual galaxies, and demonstrate that SHARDS photometry offers accurate estimates (when compared to higher resolution spectroscopy) if corrected for the limited spectral resolution. We also explore the relationships linking stellar mass, rest-frame (U-V) colour, and \Dn index for the sources in the sample, and compare them to those found in local galaxies. Using a novel approach, we combine information from the restframe (U-V) colour and \Dn to obtain an estimate of the extinction affecting the stellar population that is relatively insensitive to degeneracies with age, metallicity, and SFH. 

The paper is structured as follows: \S2 describes the sample selection and basic sample properties. \S3 describes the method used to measure the strength of the 4000 \AA{} break in the SHARDS photometry, its uncertainties, and the calibration performed using higher resolution spectra. \S4 presents the main results and \S5 discusses the implications for the downsizing scenario. Throughout this paper, we use a cosmology with H$_0$ = 70 km s$^{-1}$ Mpc$^{-1}$, $\Omega_M$ = 0.3, and $\Omega_\Lambda$ = 0.7. All magnitudes refer to the AB system.

\section{Data}\label{sample}

In this paper we analyse a stellar mass selected sample. 
Our parent sample is the GOODS-North catalog from Rainbow (PG08), with IRAC 3.6 \um as the selection band. The IRAC catalog is estimated to be 75\% (90\%) complete at $S_{\rm{3.6}}$ = 1.6 (5.0) \uuJy.

We restrict the sample to galaxies within the 141 arcmin$^2$ area surveyed by SHARDS, and also apply stellar mass and redshift constraints to ensure a highly complete sample with reliable measurements of the 4000 \AA{} break in the SHARDS SED (see \S\ref{sampleselection} for the details on the sample selection).

The SHARDS filterset consists of 24 contiguous medium-band optical filters. All of them have FWHM = 170 \AA{} except the two reddest ones, which have FWHM = 350 \AA. As of May 2013 SHARDS observations are still ongoing, and the current spectral coverage is limited to 16 filters spanning the 619--883 nm range.
 
The reduction and calibration procedures applied to SHARDS data are described in great detail in \citet{Perez-Gonzalez13}. We produced source catalogs by merging the lists of sources detected in individual filters, and then forcing measurements at all bands. 
The FWHM of the point spread function (PSF) of the images varies between 0.78'' and 1.05'', depending on the observing conditions. 

Photometry was extracted using apertures of radii $r$=0.88''. This is large enough for PSF variations to be negligible: the median colour difference between photometry in apertures with radii 0.88'' and 1'' is $<$0.03 magnitudes for any pair of images. 
Using a larger aperture (such as 1'' radius) increases the risk of contamination from nearby sources, while the SNR is largely unaffected. The median SNR increase going from 0.88'' to 1'' is 4\% for bright sources (m$<$24), and just 0.33\% for 25$<$m$<$26 sources.

We found the closest SHARDS counterpart to each Rainbow source using a 1'' search radius. However, the typical distance is much lower (median: 0.27'', 95th percentile: 0.6''). Some Rainbow sources do not have a SHARDS counterpart within 1''. These turn out to be mostly faint sources in the vicinity of a bright star. Since they represent a small fraction ($<$5\%) of the total, the results of this work will not be affected.

\subsection{Redshifts}\label{redshifts}

The GOOODS-N field has been thoroughly explored by redshift surveys, resulting in a large fraction of sources having spectroscopic redshifts down to very faint magnitudes.
We use the spectroscopic redshift catalog from the Rainbow database, which contains reliable spectroscopic redshifts for 80\% of the sources brighter than $R$=24. However, since we aim at a mass-complete sample, photometric redshifts are required to avoid underepresentation of the faintest sources within our mass and redshift limits.

The photometric redshifts of Rainbow sources are described in PG08. They use standard template-fitting through $\chi^2$ minimisation with a large set of SPS models covering a range of star formation histories and extinction.
A preliminary catalog of photometric redshifts using both broadband and SHARDS photometry is also available for sources brighter than K=24 (Ferreras et al., in preparation). It uses a template-matching technique, comparing both broadband photometry (from $U$ to Spitzer/IRAC 3.6\uu) and the medium band SHARDS data with a grid of 2000 synthetic templates built from the models of \citet{Bruzual03}. The models explore a wide range of star formation histories, including dust and emission lines. A comparison with the available spectroscopic redshifts in GOODS-N gives an accuracy $\Delta$($z$)/(1+$z$)=0.02 (median) for the whole SHARDS dataset down to $K$$<$24 in the redshift interval of our working sample (with a 1-$\sigma$ scatter about the median of just $0.03$), in contrast to $\sim$0.05 for the photometric redshifts from broadband data alone.

\subsection{Stellar masses}\label{masses}

\begin{figure}
\begin{center}
\includegraphics[width=8.5cm]{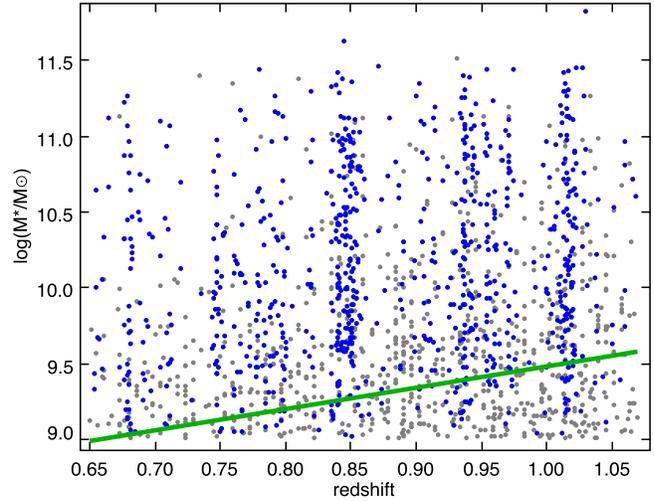}
\end{center}
\caption[]{Stellar masses versus redshift for sources with spectroscopic (blue) and photometric (grey) redshifts studied in this work. The solid green line represents the 75\% completeness limit of Rainbow for a passively evolving stellar population (PG08).\label{mass-z}}
\end{figure}

The procedure used to estimate stellar masses is described in great detail by PG08. Very briefly,
they used a maximum likelihood estimator to find the SPS model that best fits all the available photometric data points for wavelengths $<$4\um (rest-frame). The stellar emission in the models was taken from the PEGASE code \citep{Fioc97} assuming a \citet{Salpeter55} IMF with 0.1$<$M$_*$/M$_\odot$$<$100, and star formation histories (SFH) described by a declining exponential law with timescale $\tau$ and age $t$ [i.e, SFR($t$) $\propto$ $e^{-t/\tau}$].
Stellar masses calculated in this way are considered to be accurate to within a factor of 2--3.
The 75\% completeness limit of Rainbow for a passively evolving population is at $\log$(M$_*$/M$_\odot$)=9.0 at $z$=0.65 and $\log$(M$_*$/M$_\odot$)=9.5 at $z$=1.07 (PG08, see also Figure \ref{mass-z}). However, most M$_*$$\sim$10$^{9.0}$ M$_\odot$ galaxies at these redshifts are actively forming stars and are therefore significantly brighter than this limit. Since only 7\% of galaxies in the 9.0--9.5 mass interval are red [(U-V)$_r$$>$1.4], we estimate 97--98\% completeness in mass at M$_*$$\sim$10$^{9.0}$\nolinebreak M$_\odot$.

\subsection{Restframe colours}

Rest-frame optical colours are frequently used to distinguish young from old galaxies in photometric surveys \citep[e.g.][]{Strateva01,Blanton03,Baldry04}. This is because older stellar populations usually have redder optical SEDs compared to younger ones (albeit extinction by dust can also redden the optical SED of a young galaxy, see Figure \ref{models}).

We compute rest-frame magnitudes in several UV and optical broadband filters by convolving  the SPS model that best fits the broadband SED of each source with the filter transmission curve, as described in PG08. The broad spectral coverage of Rainbow photometry implies that rest-frame photometry is interpolated between observed bands. Owing to the accurate photometric redshifts, the uncertainty in rest-frame colours is comparable to the uncertainty in observed colours, $\sim$0.1 magnitudes.

In this work we will concentrate specifically on the rest-frame U-V colour, (U-V)$_r$, which is particularly sensitive to the age and metallicity of the stellar population because it straddles the 4000 \AA{} break, and has been widely used in the literature \citep[e.g.][]{Sandage78,Bower92,Bell04,Silverman09,Cardamone10}.

\subsection{Selection of a mass-complete sample}\label{sampleselection}

The current wavelength coverage of the SHARDS photometry allows for measurement of the 4000 \AA{} break in sources at redshift 0.65$\le$$z$$\le$1.07. Therefore, we selected from Rainbow only the sources with redshifts (either spectroscopic or photometric) within these limits. To ensure high completeness at $z$$\sim$1 we also restrict the sample to M$_*$$>$10$^{9.0}$ M$_\odot$ sources.
In summary, the selection criteria are as follows: 

i) Rainbow source in the GOODS-North field with a 3.6\um detection

ii) within the area covered by the SHARDS survey 

iii) 0.65 $\le$ $z$ $\le$ 1.07 (either photometric or spectroscopic)

iv) $\log$(M$_*$/M$_\odot$) $\ge$ 9.0

There are 1644 Rainbow/SHARDS sources meeting these criteria. Table \ref{general-table} summarises the overall properties of the sample. 837 [51\%] of the sources in the sample have reliable spectroscopic redshifts (hereafter, the \zspec subsample) and the remaining 807 [49\%] depend on a photometric redshift estimate (hereafter, the \zphot subsample). 562 of the \zphot sources have redshift estimates from the combined SHARDS and broadband SEDs, while the remaining 245 have redshifts from broadband alone.

\begin{figure}
\begin{center}
\includegraphics[width=8.5cm]{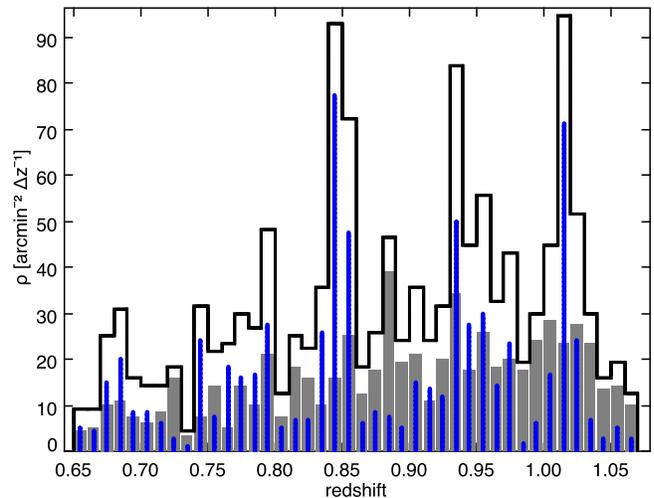}
\end{center}
\caption[]{Redshift distribution for the sources in the sample. The black solid line represents the total density of sources per unit of redshift, while grey bars and blue spikes represent the populations with photometric and spectroscopic redshift, respectively.
\label{z-distr}}
\end{figure}

The redshift distribution (Figure \ref{z-distr}) is rather inhomogeneous due to the small size of the survey area, which makes it sensitive to large scale structure. However, we consider the impact of environment on the results of this work to be negligible (see \S\ref{trends-with-z}).

Stellar masses could be overestimated in sources where emission from an AGN increases the observed flux in the near- and mid-infrared bands. A cross-correlation with the Chandra 2 Ms catalog from \citet{Alexander03} finds 56 X-ray sources in the mass-selected sample, but 54 out of the 56 show a clear 1.6 \um bump, suggesting the stellar population dominates the rest-frame near-infrared emission \citep[Hern\'an-Caballero et al., in preparation; see also][]{Alonso-Herrero08}.

\begin{figure}
\begin{center}
\includegraphics[width=8.5cm]{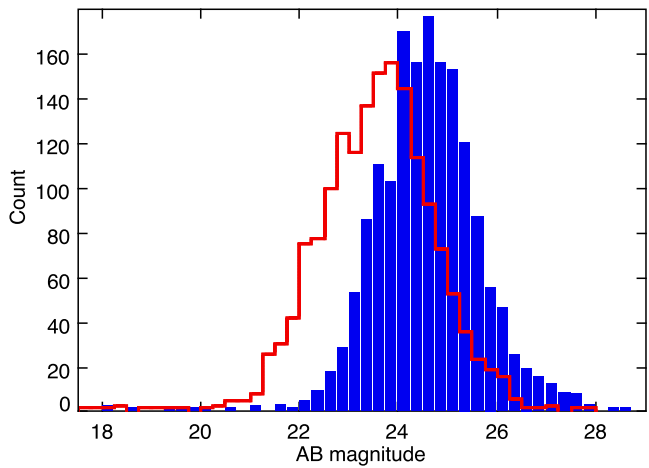}
\end{center}
\caption[]{Distribution of optical magnitudes in the SHARDS bands F636W17 (blue) and F857W15 (red) for sources in the mass-selected sample.\label{mag-histog}}
\end{figure}

Figure \ref{mag-histog} shows the distribution of the optical magnitudes, measured in apertures of radius 0.88 arcseconds, for two SHARDS bands close to the blue and red ends of the SHARDS spectral range, namely F636W17 and F857W17.
The magnitude distribution peaks at $m$=25 (24) in the F636W17 (F857W17) bands. Note that the reduction in the number counts for fainter magnitudes is due to the redshift and mass constraints of the sample selection and not the depth of the SHARDS data, which reaches $m$=26.5 (3$\sigma$) in all bands \citep{Perez-Gonzalez13}.

The typical F636W17-F857W17 color index is $\sim$1 magnitude. This implies somewhat higher signal to noise ratios (SNR) at longer wavelengths for most sources. For example, the fraction of sources with SNR$>$5(10) increases from 93.4\%(85\%) in the F636W17 band to 98.5\%(93.7\%) in the F857W17 band.

\section{Measurement of the 4000 \AA{} index}\label{sec-measure-4000}

The break in the stellar continuum at 4000 \AA{} is the strongest discontinuity in the optical spectrum of galaxies older than $\sim$1 Gyr, and arises because of the accumulation of a large number of spectral lines in a narrow wavelength region. 
The main contribution to opacity comes from ionised metals. These are highly ionised in hot stars, effectively decreasing the opacity, and therefore O and B type stars have much weaker breaks compared to later spectral types \citep{Bruzual83}. Since hot (massive) stars have short lives, the strength of the 4000\nolinebreak{} \AA{} break for a single stellar population (SSP) increases with its age. There is also a dependency with the metallicity, but it only becomes significant for old stellar populations (see Figure \ref{models}).
In the integrated spectrum of a galaxy, emission from hot young stars can easily outshine the old population and produce a weak 4000\nolinebreak{} \AA{} break even if they only represent a small fraction of the total stellar mass in the galaxy. Therefore, the 4000 \AA{} break provides a light-weighted characteristic age for the galaxy, which is close to the mass-weighted age only for passively evolving galaxies.

The strength of the 4000 \AA{} break is commonly measured using one of two definitions: the ``regular'' one, D(4000), measures the ratio between average spectral flux densities (f$_\nu$) in the rest-frame bands [4050,4250] and [3750,3950] \AA{} \citep{Bruzual83, Hamilton85}, while the ``narrow'' index, \Dnu, uses the ratio between the bands at [4000,4100] and [3850,3950] \AA{} \citep{Balogh99}. The narrow index has the advantage of being less sensitive to extinction and is the preferred choice in most recent studies \citep[e.g.][]{Kauffmann03b,Martin07,Silverman09,Gobat12}.

To calculate the average f$_\nu$ in each of these bands, we shift the SHARDS SED to the rest-frame of the source, and perform a linear interpolation of the spectrum between adjacent photometric points. 

Visual inspection of all the individual SHARDS SEDs reveals 66 sources with features close to 4000 \AA{} rest-frame that do not correspond with a 4000 \AA{} break. Only 8 out of these 66 sources have spectroscopic redshifts. This suggests many of them could have larger than usual redshift errors. We cautiously put aside these sources in the analysis of the 4000 \AA{} break strengths. Since they represent just 4\% of the sample, our results will not be significantly affected.

\begin{figure}
\begin{center}\hspace{-0.3cm}
\includegraphics[width=8.5cm]{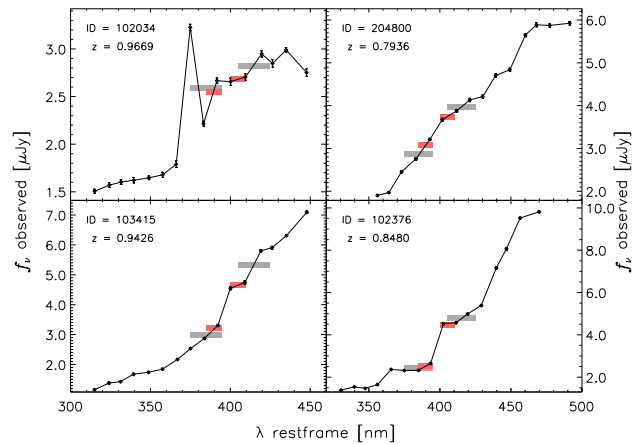}
\end{center}
\caption[]{Rest-frame 300--450 nm SHARDS SED of four representative sources from the sample. Error bars represent the 1-$\sigma$ uncertainty due to noise variance in the flux; the uncertainty in the photometric zero-point (0.05--0.07 magnitudes) is not included. The grey and red bars represent the bands that define the D(4000) and \Dn indices, respectively. The large spike at $\sim$375 nm is due to the \oii 3727 \AA{} emission line entering one of the SHARDS filters.\label{actualdata}}
\end{figure}

There are three main sources of uncertainty affecting the 4000\nolinebreak{} \AA{} break measurements: photometric errors, interpolation errors, and redshift uncertainty. We discuss them in the following.

Photometric errors are lower than 0.1 magnitudes for 83\% of sources in the F636W17 band, and 93\% in the F857W17 band. Nevertheless, the correlation between stellar mass and optical magnitude implies that the photometric errors increase with decreasing stellar masses. For galaxies in the lowest bin of mass considered here (9.0 $\le$ $\log$(M$_*$/M$_\odot$) $<$ 9.5), 50\% (15\%) of sources have flux uncertainties larger than 0.1 magnitudes in the F636W17 (F857W17) filter.

The interpolation error is the error in the determination of the index due to the limited spectral resolution of the SHARDS SED. Given the wavelength resolution of the SHARDS SEDs, 1 or 2 photometric points are within the limits of each rest-frame band defining the \Dn index, and 2 or 3 in the case of the D(4000) index. Figure \ref{actualdata} shows the SHARDS SED and integration bands for a few typical sources. 

Compared to a high resolution spectrum, the degradation in the spectral resolution introduced by the convolution with the transmission profiles of the SHARDS filterset reduces the contrast of the 4000 \AA{} break, and thus it tends to decrease the measurement of the index. This decrement is larger for \Dn compared to D(4000). For sources with intense emission in the \oii 3727 \AA{} line, the blue band defining the D(4000) index can be contaminated (see top left panel in Figure \ref{actualdata}), reducing the apparent strength of the break. To mitigate this issue, in sources with clear \oii emission we substitute the observed flux in the affected band by a log-linear interpolation of the adjacent ones.

To quantify the influence of the photometric and interpolation errors in the 4000 \AA{} break measurement, we obtained synthetic SHARDS photometry on a large sample of optical spectra drawn from the zCOSMOS \citep{Lilly07} Data Release 2. We compared D(4000) and \Dn values measured in the synthetic photometry with those from the full-resolution spectrum, and obtained a calibration that converts raw D(4000) values into interpolation  corrected \Dn values. Our procedure for obtaining the synthetic photometry and the calibration is similar to that described by \citet{Kriek11}, but it uses real spectra as reference instead of SPS models, and is applied to individual galaxies instead of average SEDs. 
Further details on the calibration method are provided in the appendix at the end of this paper.

The systematic error in \Dn introduced by the calibration model is at most $\sim$3\%,
while the 1-$\sigma$ dispersion of \Dn values obtained from the synthetic photometry for a given \Dn is $\sim$5\% and $\sim$8\% for photometric uncertainties of $\Delta$m = 0.01 and 0.1 magnitudes, respectively.

The total uncertainty in \Dn measurements of SHARDS sources due to the combination of photometric, interpolation, and calibration errors is estimated to be $\sim$6\% and $\sim$9.5\% of the measured value at m=22.5 and 25.5, respectively.

The redshift uncertainty is significant only for sources that rely on a photometric redshift estimate, that is, about half of the sample (see \S\ref{sample}).
Comparison of photometric and spectroscopic redshifts for the other half indicates that the distribution of relative errors $\Delta$($z$)/(1+$z$) in the photometric redshifts has median 0.02 and standard deviation 0.03.

\begin{figure}
\begin{center}
\includegraphics[width=8.5cm]{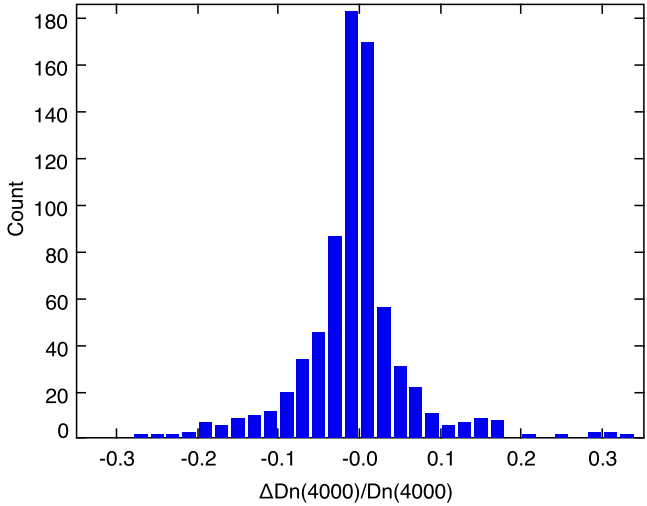}
\end{center}
\caption[]{Distribution of the relative difference between \Dn values obtained using the photometric and spectroscopic redshift, $\Delta$\Dnu/\Dnu, for the sources where both redshifts are within the 0.65$\le$$z$$\le$1.07 range. 80\% of sources have $\vert$$\Delta$\Dnu$\vert$/\Dnu$<$0.05, and 93\% have $\vert$$\Delta$\Dnu$\vert$/\Dnu$<$0.1.\label{delta-Dn4000}}
\end{figure}

Redshift uncertainties of this order are still sufficiently large to move in a few cases \textit{both} bands defining the index to the same side of the 4000 \AA{} break, potentially compromising the measurement. Nevertheless, comparison of \Dn values obtained using both the photometric and spectroscopic redshifts for the \zspec subsample indicates that they are within 5\% (10\%) of each other in 80\% (93\%) of cases (Figure \ref{delta-Dn4000}). Since the \zspec subsample is biased in favour of sources with emission lines, the impact of photometric redshift errors in \Dn values for the \zphot subsample should be even lower.

In summary, the accuracy of \Dn measurements for the mass-selected sample of SHARDS galaxies is expected to range from $\sim$5\% in massive galaxies with spectroscopic redshifts to $\sim$10--15\% in low mass galaxies with photometric redshifts.

As a sanity check, we have obtained direct measurements of the D(4000) index in a sample of 56 high SNR HST/ACS spectra from the PEARS project\footnote{Downloaded from the database at http://archive.stsci.edu/prepds/pears/. See also \citet{Ferreras09}.} which have no signs of contamination from other PEARS sources in their slitless grism spectra. Comparison of D(4000) from the PEARS spectrum and the SHARDS SED shows that both estimates are compatible within their uncertainties (the 1-$\sigma$ dispersion in $\Delta$D(4000)/D(4000) is 0.081).

\section{Results}

\subsection{Color-magnitude diagram}\label{sect-CMD}

\begin{figure}
\begin{center}
\includegraphics[width=8.5cm]{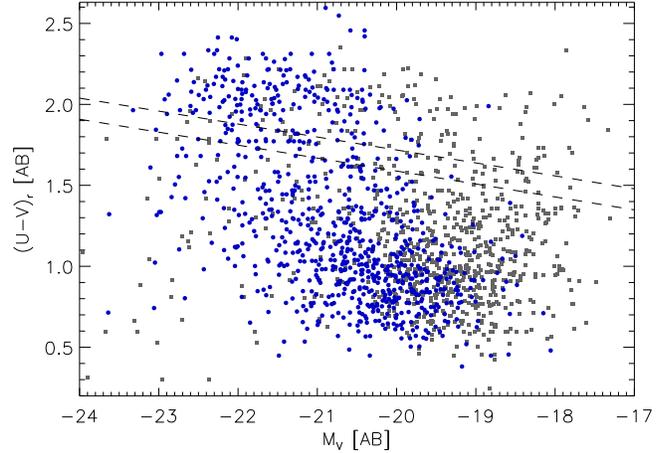}
\end{center}
\caption[]{Rest-frame (U-V) colour versus absolute V-band magnitude for the sources with spectroscopic (blue) and photometric (grey) redshifts. The upper and lower dashed lines represent the limit of the red sequence as defined by \citet{Bell04} for redshifts $z$=0.65 and 1.07, respectively.\label{CMD}}
\end{figure}

It is well known that the colour distribution of galaxies depends on both luminosity and redshift \citep[e.g.][]{Bell04}. 
For a fixed absolute magnitude, (U-V)$_r$ decreases with increasing redshift, while at a given redshift (U-V)$_r$ increases with the luminosity. 

Using a large sample of $\sim$25000 sources from the COMBO-17 survey \citep{Wolf03} in the redshift range 0.2$<$$z$$\le$1.1, \citet{Bell04} found that the average (U-V)$_r$ for red sequence galaxies with absolute V-band magnitude M$_V$=-20 is:
\begin{equation}
\langle (U-V) \rangle_{M_V = -20} =  2.17 - 0.31 z
\end{equation}
where we have included the 0.77 magnitudes difference between U-V colours in the AB and Vega systems.
Given the limited redshift range covered by our sample, the evolution of the red sequence is expected to be small ($\sim$0.13 magnitudes between $z$=0.65 and $z$=1.07, comparable to the uncertainty in the (U-V)$_r$ colour of individual galaxies). 

The slope of the red sequence, $\frac{d\langle(U-V)\rangle}{dM_V}$, is between -0.05 and 0.1 both in local cluster and field galaxies \citep{Schweizer92,Bower92,Terlevich01,Baldry04}. For consistency, we assume the value -0.08 used by \citet{Bell04} in their analysis.

\begin{figure}
\begin{center}
\includegraphics[width=8.5cm]{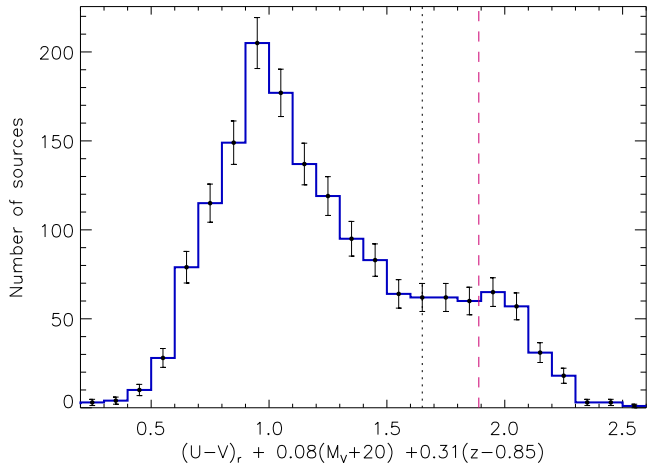}
\end{center}
\caption[]{Distribution of rest-frame (U-V) colour corrected for the dependency with absolute magnitude and redshift as defined by \citet{Bell04}. The correction is 0.0 for a M$_V$=-20 galaxy at $z$=0.85 and reaches $\pm$0.3 in extreme cases. Errors calculated as the squared root of the number of sources in each bin. The red dashed line represents the mean U-V colour of red sequence galaxies from \citet{Bell04}, and the dotted line marks their definition of the red sequence limit.\label{UVseq-histog}}
\end{figure}

Figure \ref{CMD} shows the (U-V)$_r$ versus M$_V$ color-magnitude diagram (CMD) of our sample. The dashed lines represent the red sequence cuts for redshifts 0.65 and 1.07, using the definition from \citet{Bell04}:
\begin{equation}
(U-V)_r > 1.92 - 0.31 z - 0.08 (M_V + 20)
\end{equation}

With this definition, the red sequence contains 350 of the 1669 sources in the sample.

The bimodality in the distribution of the (U-V) colour is more clearly observed if we correct for the dispersion introduced by the dependency with $z$ and M$_V$. We define the equivalent (U-V) colour at redshift $z$=0.85 and M$_V$=-20 as: 
\begin{equation}
(U-V)_0 = (U-V)_r + 0.31(z-0.85) + 0.08(M_V+20).
\end{equation}

Figure \ref{UVseq-histog} shows the distribution of (U-V)$_0$ for the sample. The distribution has a main peak at (U-V)$_0$=1.0 corresponding to the blue cloud, and a secondary, broader one centered at (U-V)$_0$$\sim$1.9. The minimum between them is at (U-V)$_0$$\sim$1.55, close to the red sequence cut which is at (U-V)$_0$=1.65.

\subsection{Trends with stellar mass}\label{trends-with-mass}

\begin{figure*}
\begin{center}
\includegraphics[width=17.7cm]{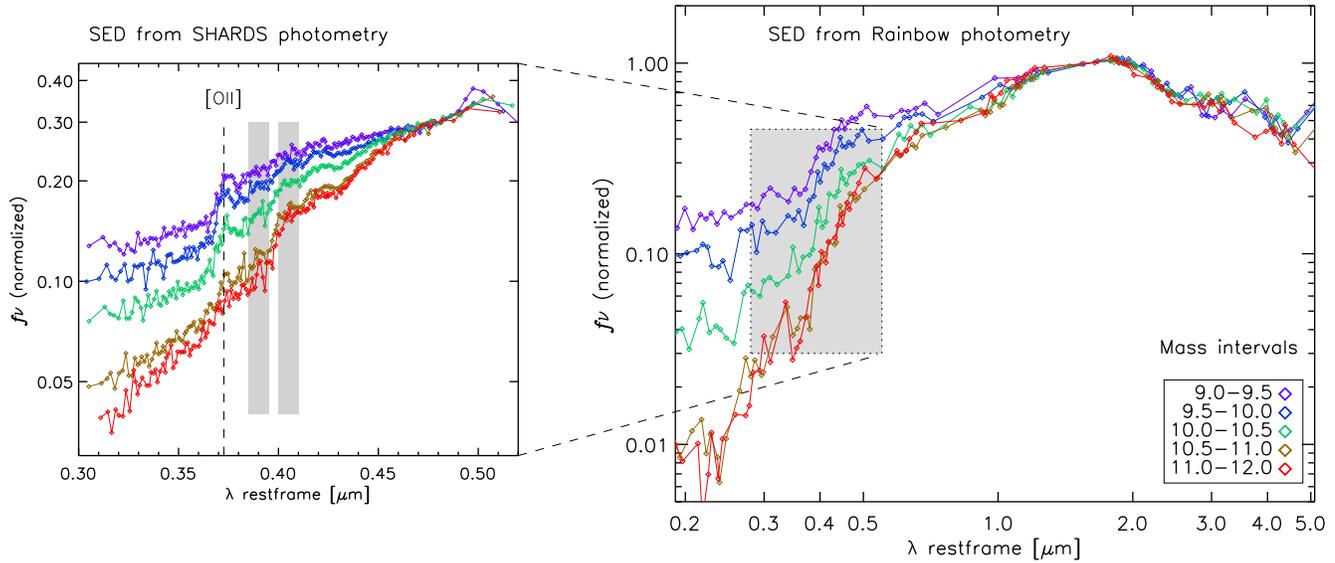}
\end{center}
\caption[]{The mass dependency of the medium band (left) and broadband (right) SEDs of SHARDS galaxies. Each SED is the average of all the sources in the sample within a given stellar mass interval, normalised at $\lambda$ = 0.5 \um for the SHARDS SED and 1.6\um for the Rainbow SED. The two parallel stripes in the left plot represent the bands defining the \Dn index, while the dashed line marks the position of the \oii 3727 \AA{} emission line.\label{compareSEDs}}
\end{figure*}

Figure \ref{compareSEDs} compares the stacked SED for the sources contained in 5 intervals of log(M$_*$/M$_\odot$), from 9.0--9.5 to 11.0--12.0. SEDs from the broadband photometry are normalised at the 1.6 \um peak of the stellar emission. The NIR SED ($\lambda$=1--5\uu) of these sources shows little dependency with the stellar mass, unlike the rest-frame optical range, where there is a clear trend of increasingly red SEDs at higher stellar masses. This is the expected result if the more massive galaxies contain on average older stellar populations, but it could also be produced by increased obscuration or higher metallicity in more massive galaxies. Further insight is provided by the stacked SHARDS SEDs (left panel in Figure \ref{compareSEDs}). The SEDs for the three lower mass intervals show strong emission in the \oii 3727 \AA{} line and a weak 4000 \AA{} break, both features indicative of a young stellar population. On the other hand, the SEDs of the two higher mass intervals shown no significant \oii emission and a stronger 4000 \AA{} break, hinting at older average stellar ages. This suggests that stellar age and not obscuration is the main driver of the correlation between optical colours and stellar mass (see also \S\ref{UV-Dn4000-subsection}).

\begin{figure}
\begin{center}
\includegraphics[width=8.5cm]{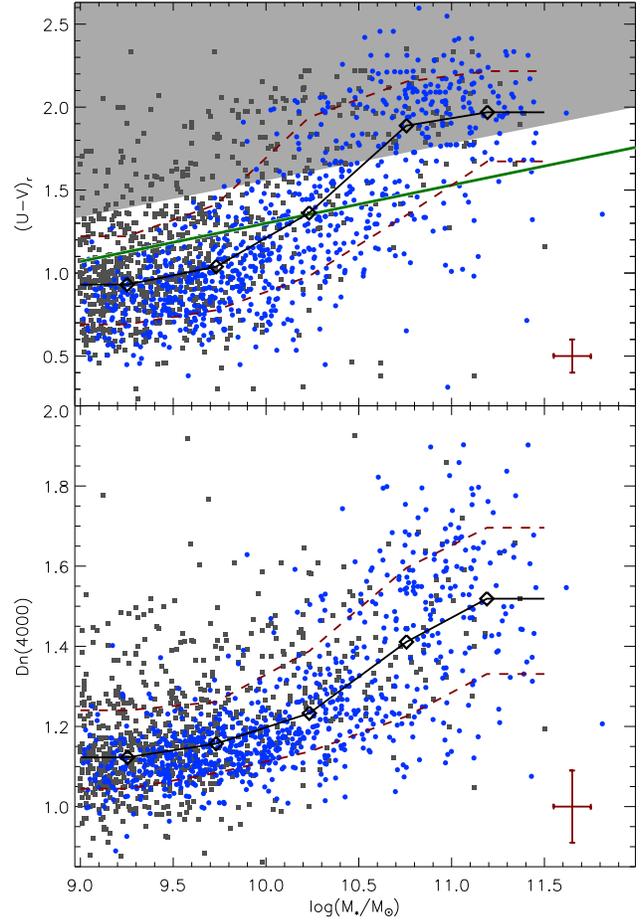}
\end{center}
\caption[]{Rest-frame observed U-V colour (not corrected for extinction; top panel) and \Dn index (bottom panel) as a function of the stellar mass. Blue dots and grey squares represent sources with spectroscopic and photometric redshifts, respectively. Typical uncertainties are shown in the lower right corner. The open diamonds connected by a solid line represent median values in bins of mass 0.5 dex wide, while dashed lines indicate the 16$^{th}$ and 84$^{th}$ percentiles. The shaded area in the top panel represent the location of the red sequence above the cut defined by \citet{Cardamone10} for the redshift range 0.8$<$$z$$<$1.2 and the green dashed line represents their location for the green valley in the same redshift range.
\label{Dn4000-UV-mass}}
\end{figure}

When considering individual sources, both the \Dn index and the (U-V)$_r$ colour correlate with the stellar mass, albeit with a high dispersion (see Figure \ref{Dn4000-UV-mass}).
In the (U-V)$_r$ vs M$_*$ diagram there seems to be a smooth transition between the blue cloud and the red sequence, with galaxies in the 10.0$<$$\log$(M$_*$/M$_\odot$)$<$10.5 range placed in most cases within or near the region corresponding to the green valley at these redshifts.

\begin{figure}
\begin{center}
\includegraphics[width=8.5cm]{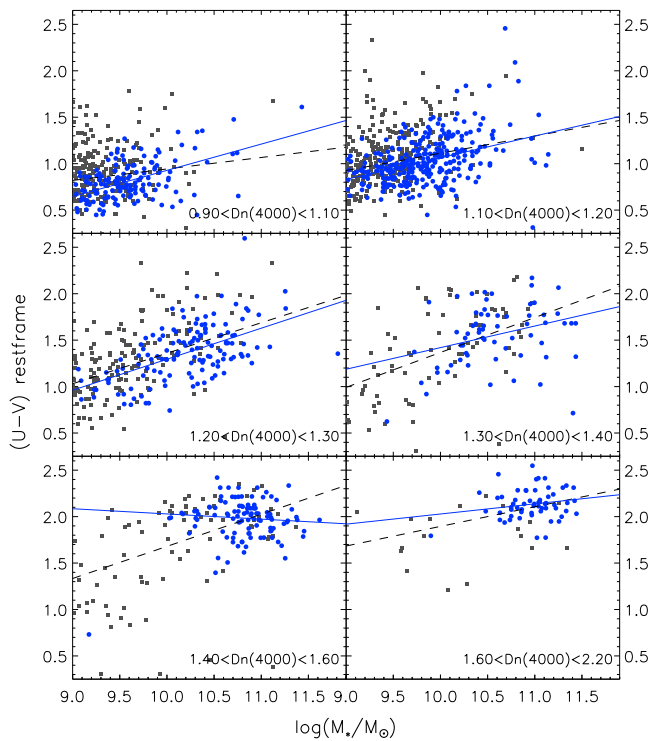}
\end{center}
\caption[]{Rest-frame observed (U-V) colour (not corrected for extinction) versus stellar mass for several intervals of \Dnu. Blue dots and grey squares represent sources with spectroscopic and photometric redshifts, respectively. The dashed line represents the best fitting linear relation considering both \zspec and \zphot sources, while the solid line represents the best linear fit when considering only sources with spectroscopic redshift.\label{UV-mass-by-Dn4000}}
\end{figure}

We note that the correlation between (U-V)$_r$ and log(M$_*$) is somewhat stronger if considering only the \zspec subsample. ($r$ = 0.76 vs $r$ = 0.65 for the entire sample). This is because red (U-V)$_r$ colours at low masses are found almost exclusively in \zphot sources.
This does not necessarily imply poor accuracy in the (U-V)$_r$ measurements of \zphot sources, since redshift uncertainties are too low compared to the width of the U and V filters to affect significantly the (U-V)$_r$ color. Rather, it illustrates the bias of spectroscopic samples (which usually only detect low mass galaxies if they show prominent emission lines) relatively to a mass-limited sample. In particular, the \zphot subsample shows what looks like a continuation of the red sequence (i.e., redder (U-V)$_r$ colours) down to the mass limit of the sample. As Figure \ref{UV-mass-by-Dn4000} demonstrates, many of the sources in the extended red sequence have low \Dn values consistent with a young but obscured stellar population. However, some of them show larger \Dn values, consistent with a transitioning or quiescent galaxy. 

The distribution of \Dn versus M$_*$ shows the same general trend of increasing \Dn values at higher masses. Average \Dn values remain nearly constant at \Dnu=1.1--1.2 up to M$_*$$\sim$10$^{10.25}$M$_\odot$, but they increase steadily for more massive galaxies (see also Table \ref{general-table}).

While the dispersion in (U-V)$_r$ for a given mass is roughly constant in the entire mass range, maybe even decreasing at the very high end, for \Dn the dispersion increases significantly with M$_*$. This different behaviour is probably due to the higher influence of extinction on the (U-V)$_r$ colour (see \S\ref{UV-Dn4000-subsection}).

An old massive galaxy with M$_*$=10$^{11}$M$_\odot$ and \Dnu=1.8, if scaled down, would be detectable by SHARDS down to at least M$_*$=10$^{9.5}$M$_\odot$ at $z$$\sim$1. Given that, it is noteworthy the rarity of old (\Dnu$>$1.6) galaxies with M$_*$$<$10$^{10.5}$ M$_\odot$.
 
The region defined by M$_*$$<$10$^{10}$ M$_\odot$ and \Dnu$>$1.3 has a very small fraction of \zspec sources. Sources with photometric redshift in this region are mainly the galaxies with red (U-V)$_r$ in the extended red sequence, but there are also a few young starbursts where \Dn is grossly overestimated due to the \oii 3727 \AA{} emission line entering the red band defining the \Dn index because of an overestimated photometric redshift. The latter can be identified because of their inconsistently low (U-V)$_r$ values (see \S\ref{UV-Dn4000-subsection}).

Although the correlation of (U-V)$_r$ and \Dn with the stellar mass is mainly driven by an increase in the average stellar age in more massive galaxies, the effects of metallicity and extinction cannot be dismissed. In fact, it is well known that metallicity is tightly correlated with the stellar mass at least up to $z$$\sim$1 \citep[the so-called ZM relation; e.g.][]{Tremonti04, Savaglio05, Lamareille09, Yabe12}.

Since \Dn is much less sensitive to extinction compared to (U-V)$_r$ and nearly independent of metallicity for stellar populations younger than $\sim$1 Gyr (see Figure \ref{models}), we can test the impact of extinction in the relationship between the (U-V)$_r$ colour and the stellar mass by selecting subsamples of sources within narrow intervals of \Dnu.

Figure \ref{UV-mass-by-Dn4000} shows that there is a general trend of increasing (U-V)$_r$ with M$_*$ at a fixed \Dnu, albeit with high dispersion. This suggests that for a given stellar age, extinction tends to be higher in the more massive galaxies. Higher extinction in massive but young galaxies would explain our finding of a lower dispersion in the (U-V)$_r$ values of massive galaxies compared to their \Dn values.
For galaxies in the two higher \Dn bins (bottom row in Figure \ref{UV-mass-by-Dn4000}) the correlation seems to flatten or even reverse, at least for the \zspec subsample. However, the uncertainty in the slope is higher than in younger subsamples because there are very few low mass galaxies with high \Dn values, and nearly all of them have photometric redshifts. In addition, we caution that in the older galaxies (those with \Dnu$>$1.6) the effect of metallicity in \Dn and (U-V)$_r$ is no longer negligible. In \S\ref{ssec-Av-mass} we revisit the mass-extinction relationship with quantitative estimates of A$_V$.

\subsection{Dependency with redshift}\label{trends-with-z}

The redshift range covered by the sample equals 2 Gyr of cosmic evolution. If the so called downsizing in hierarchical galaxy formation is significant from z=1.07 to z=0.65, 
then we should be able to observe variations in the distribution of the stellar age indicators between the sources with lower and higher redshift values, at least for the
mass bins where the bulk of the transition from actively star-forming to quiescent is believed to be taking place at these redshifts.

\begin{figure}
\begin{center}
\includegraphics[width=8.5cm]{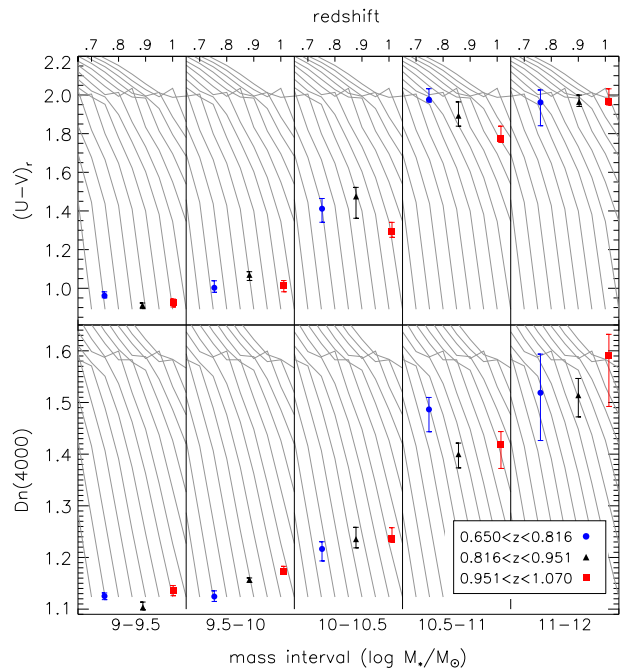}
\end{center}
\caption[]{Dependency with redshift of the median values of the rest-frame (U-V) colour (top) and the 4000 \AA{} index (bottom) for several mass ranges. Solid symbols represent the median values in three redshift intervals for each mass range, with error bars indicating the 67\% confidence interval calculated with bootstrap resampling. The redshift scale is shown at the top of the plot. The grey solid lines represent tracks for passive evolution of a SSP with moderate extinction (A$_V$=0.5). All the tracks represent the redshift dependency of the rest-frame (U-V) colour and \Dn index of a SSP with different formation redshifts.
\label{z-evolution}}
\end{figure}

To look for hints of this evolution, we separate the sources into three redshift intervals representing comparable comoving volumes: [0.65,0.816), [0.816,0.951), and [0.951,1.07]. 
Figure \ref{z-evolution} shows the median values of \Dn and (U-V)$_r$ for the three subsamples, separated into 5 mass bins.

The U-V colour shows a small but significant evolution with redshift, being $\sim$0.1--0.2 magnitudes redder at $z$$\sim$0.7 compared to $z$$\sim$1. The redshift evolution seems to increase with the stellar mass, except maybe in the highest mass interval (M$_*$/M$_\odot$$>$10$^{11}$) where the larger error bars obfuscate the trend.
In the diagram for \Dn the evolution is much less pronounced, with values being compatible within their uncertainties with a flat slope. 

For comparison, tracks corresponding to passive evolution are shown as grey lines in Figure \ref{z-evolution}. The slope of these tracks is much steeper compared to the redshift evolution observed in the sample, even in the 10.5--11.0 mass range, indicating that only a small fraction of the galaxies would be evolving passively.
Observing significant evolution in the (U-V)$_r$ colour but not in \Dn suggests that the redshift dependency of reddening in (U-V)$_r$ might in fact be related to increased obscuration or higher metallicity at lower redshifts, instead of older average light-weighted stellar ages.

In such a small field, clustering could contribute a fraction of the scatter observed. It is evident from Figure \ref{z-distr} that the volume sampled contains several overdensities and large voids. Nevertheless, several works have demonstrated that the environment dependency of galaxy colours is weak, at least for early type galaxies \citep[e.g.][]{Balogh04,Hogg04,Bernardi06}. 
In the zCOSMOS survey, \citet{Cucciati10} found that after accounting for the mass dependency in galaxy colours, the colour-density relation of M$_*$$>$10$^{10.7}$ M$_\odot$ galaxies is flat up to $z$$\sim$1, while at lower masses the fraction of red galaxies at 0.1$<$$z$$<$0.5 depends on the environment. Also in zCOSMOS, \citet{Moresco10} estimate a difference in the average (U-V)$_r$ of red galaxies between underdense and overdense regions of $\sim$0.03 mag, while for \Dn this difference is $\sim$0.05. Since our three redshift bins likely contain both field and cluster galaxies, the environment contribution to the variations in the median (U-V)$_r$ and \Dn with $z$ should be significantly lower (unless blue galaxies have a much stronger environment dependency).
This implies that the trends with redshift of (U-V)$_r$ shown in Figure \ref{z-evolution} are real.
 
\subsection{The (U-V) versus \Dn diagram}\label{UV-Dn4000-subsection}

\begin{figure}
\begin{center}
\includegraphics[width=8.5cm]{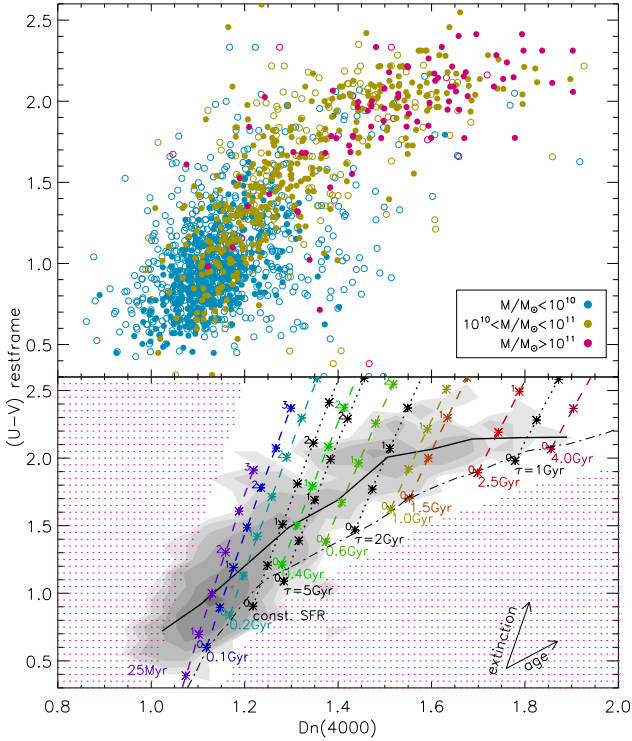}
\end{center}
\caption[]{(Top) rest-frame (U-V) color index versus \Dn for individual SHARDS galaxies. Open (solid) symbols represent sources with photometric (spectroscopic) redshift. The color coding indicates the stellar mass range for each source.
(Bottom) density countours for the distribution of sources in the \zspec subsample. The black solid line indicates their median (U-V)$_r$ in bins of \Dn of width 0.1. Dashed lines represent tracks as a function of extinction for SSP models with ages (left to right): 0.025, 0.1, 0.2, 0.4, 0.6, 1.0, 1.5, 2.5, and 4 Gyr. Dotted lines represent a 6 Gyr old galaxy with an exponentially declining SFR and e-folding times $\tau$ = 1, 2, and 5 Gyr, and also $\tau=\infty$ representing a constant SFR. All templates are computed with GALAXEV \citep{Bruzual03} and assume solar metallicity and a \citet{Salpeter55} IMF. Each tick represents a 0.5 increment in A$_V$ starting from A$_V$ = 0 at the lower end (dot-dashed line), assuming a \citet{Draine03} extinction law with Milky Way grain size distribution \citep{Weingartner01} and R$_V$ = 3.1. The dotted area represents combinations of (U-V)$_r$ and \Dn values that cannot be reproduced with SSP models for any age, metallicity, or extinction.\label{Dn4000-UV-tracks}}
\end{figure}

The correlation between values of (U-V)$_r$ and \Dn for individual galaxies is most evident when representing the former against the latter (Figure \ref{Dn4000-UV-tracks}).
The top panel represents (U-V)$_r$ versus \Dn for individual SHARDS galaxies, with color coding for the stellar mass. 
In the bottom panel, contours represent the density distribution of \zspec
sources, while the black solid line represents the median (U-V)$_r$ value for \zspec sources in bins of \Dn of width $\Delta$=0.1.
The largest concentration of sources is at (U-V)$_r$$\sim$1.0 and \Dnu$\sim$1.1, and is composed mainly of galaxies with M$_*$$<$10$^{10}$ M$_\odot$. On the other hand, the most massive galaxies cluster around a horizontal branch at (U-V)$_r$$\sim$2.1

The distribution has a knee composed by sources with (U-V)$_r$ values as high as $\sim$2 magnitudes but relatively low \Dnu$\lesssim$1.4, which implies relatively young but significantly obscured stellar populations.

We have overlaid tracks for model galaxies with different SFHs.
The SED corresponding to each SFH was computed using GALAXEV \citep{Bruzual03} with the STELIB stellar library and Padova1994 isochrones \citep{Bertelli94}, assuming solar metallicity and a \citet{Salpeter55} initial mass function (IMF). \citet{Kauffmann03a} found that at fixed metallicity and age, the variations in \Dn between STELIB and other stellar libraries are $\sim$0.05.
The expected values for (U-V)$_r$ and \Dn from each SFH model vary as a function of the extinction A$_V$. Dashed lines in Figure \ref{Dn4000-UV-tracks} represent tracks for SSP models, while dotted lines represent SFHs for galaxies 6 Gyr old with an exponentially declining SFR. 

The very steep slope of the tracks shows that, irrespective of the SFH of the galaxy, the U-V color index is much more affected by extinction than \Dnu. Nevertheless, the latter is also affected, and ignoring extinction would lead to systematically overestimated stellar ages. In example, a 0.2 Gyr old SSP with A$_V$ = 2 mag. has the same \Dn index as a 0.4 Gyr old SSP with A$_V$ = 0. Fortunately, their (U-V)$_r$ colours are different, and therefore we can combine the information provided by (U-V)$_r$ and \Dn to estimate A$_V$ and obtain de-reddened colours (see \S\ref{ssec-extinction-correction}).

The curve connecting the A$_V$=0 marks of all the SSP tracks, and the track for the youngest SSP define a region in the (U-V)$_r$ vs \Dn diagram whose values can be reproduced by the models. Outside of this region (dotted area in the bottom panel) there are no combinations of SFH, metallicity, and extinction explaining their \Dn and (U-V)$_r$ values. 

Interestingly, almost all of the sources found in this area excluded by models have photometric redshifts.
Since \Dn is much more sensitive to redshift uncertainty than (U-V)$_r$, we assume that the deviations are mainly along the X axis. Given the typical uncertainties expected for \zphot sources from the simulations with spectra ($\Delta$\Dnu/\Dn $\sim$0.10--0.15; see \S\ref{sec-measure-4000}) the dispersion observed is consistent with these sources being scattered away from the region covered by models due to random errors in their measurement of \Dnu.

There are a few outliers with very large or very small \Dnu, all of them low mass \zphot sources (open blue symbols in Figure \ref{Dn4000-UV-tracks}). 
One mechanism capable of producing large errors in \Dn from a relatively small redshift error is an emission line (in particular \oii 3727 \AA{}) entering one of the bands defining the \Dn index. If the blue (red) band is contaminated then the strength of the break will be under- (over-)estimated. Visual inspection of these outliers reveals several cases of both kinds in the sample. Since these sources are low-mass galaxies with blue (U-V)$_r$ colours, intense \oii emission from a recent or ongoing starburst is not surprising. Strong emission lines may also have a noticeable effect on the broadband photometry. However, our restframe colour measurements are largely unaffected because they are performed on the SPS model that best fits the broadband SED of each galaxy.
 
\subsection{Breaking the age-extinction degeneracy}\label{ssec-extinction-correction}

\begin{figure}
\begin{center}
\includegraphics[width=8.5cm]{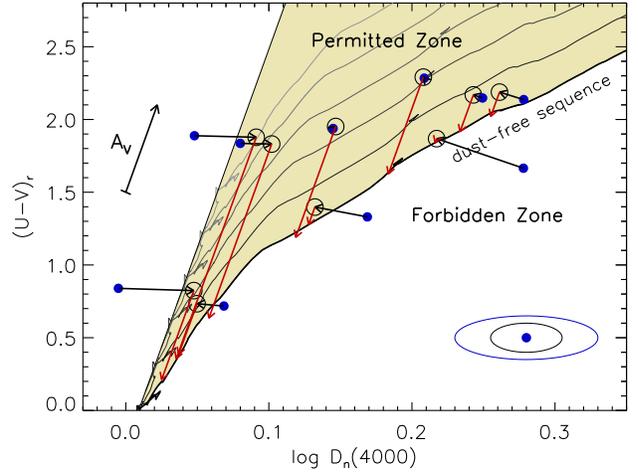}
\end{center}
\caption[]{Sketch illustrating how the extinction corrected (U-V)$_r$ and \Dn are estimated (see text). Solid symbols represent the observed (U-V)$_r$ and \Dn of a few representative sources from the sample. Open symbols represent their maximum likelihood (ML) estimates after uncertainties have been considered, and red arrows indicate the de-reddening path that takes them to the dust-free sequence containing all the extinction corrected estimates. The ellipses in the bottom right corner indicates the typical 1-$\sigma$ uncertainty in the observed values for a source in the \zphot (blue) and \zspec (black) subsamples.\label{dereddening-howto}}
\end{figure}

The (U-V)$_r$ colour and \Dn of A$_V$=0 models are tightly correlated irrespective of the metallicity (see Figure \ref{models}, right panel) or SFH (Figure \ref{Dn4000-UV-tracks}, bottom panel). We call the band occupied by A$_V$=0 models in the (U-V)$_r$ vs \Dn diagram the dust-free sequence (DFS). Variations in the age, SFH, and metallicity of the models move galaxies along the DFS, but departures from it are very small.
As a consequence, extinction is the only model parameter that can take galaxies away from the DFS. This provides us with a new way of estimating A$_V$ that does not require SED fitting and --more importantly-- is not affected by the usual degeneracy among age, extinction, and metallicity.

The increment in (U-V)$_r$ due to extinction --that is, the colour excess E(U-V)-- is proportional to A$_V$ and independent of the spectrum of the galaxy. The increment in \logDn due to extinction, $\Delta$\logDnu, is also proportional to A$_V$. While the proportionality constant for each of them is somewhat dependent on the specific extinction law assumed, the ratio E(U-V)/$\Delta$\logDn is largely insensitive to changes in the slope of the extinction curve because the bands defining the \Dn index are between the $U$ and $V$ bands.
Therefore, E(U-V) and $\Delta$\logDn are the components of a vector in the (U-V)$_r$ vs \logDn plane (the \overrightarrow{A}$_V$ vector) whose direction depends only on the effective wavelengths of the $U$ and $V$ bands and those defining the \Dn index, but is independent of any physical property of the galaxy.

Since extinction shifts the DFS in parallel to the \overrightarrow{A}$_V$ vector, we define the permitted zone (PZ) as the area of the (U-V)$_r$ vs \logDn plane that is scanned by the DFS at increasing values of A$_V$. The PZ contains all the combinations of (U-V)$_r$ vs \logDn that can be reproduced by models for any age, SFH, metallicity, or extinction.

Due to the significant uncertainty in \logDn ($\sim$0.025 for \zspec sources, $\sim$0.05 for \zphot ones) and to a lesser extent also in (U-V)$_r$ ($\sim$0.1--0.15), random errors will scatter some galaxies to a region outside of the PZ (we call it the forbidden zone, FZ).
If we assume that the errors in $\log$\Dn and (U-V)$_r$ are uncorrelated and have a normal distribution, then the probability distribution for the true values of $\log$\Dn and (U-V)$_r$ is a bidimensional gaussian function $g$($D_n$,$UV$) centered on the observed values.
If we also assume that the true values must be inside the PZ, then the maximum likelihood (ML) estimates are no longer the observed ones, but the average --weighted by the probability distribution-- of values inside the PZ. That is:
\begin{eqnarray}
\log D_n(4000)_{ML} = \iint_{PZ} g(D_n,UV)\ D_n\ dD_n\ dUV\\
(U-V)_{ML} = \iint_{PZ} g(D_n,UV)\ UV\ dD_n\ dUV
\end{eqnarray}

\noindent Figure \ref{dereddening-howto} shows that this weighted average moves sources in the FZ to the PZ, and also shifts sources close to the borders of the PZ to inner regions of it. 
To obtain de-reddened values for (U-V)$_r$ and $\log$\Dn we only need to move the position of the ML estimate along a line parallel to \overrightarrow{A}$_V$ until it intersects the DFS. The length of the segment traveled is then proportional to the implied absorption in the restframe V band, A$_V$. Assuming a \citet{Draine03} extinction law with Milky Way grain size distribution \citep{Weingartner01} and R$_V$ = 3.1, a 1 magnitude increment in A$_V$ implies E(U-V) = 0.6 and $\Delta$\logDn = 0.022.

\begin{figure}
\begin{center}
\includegraphics[width=8.5cm]{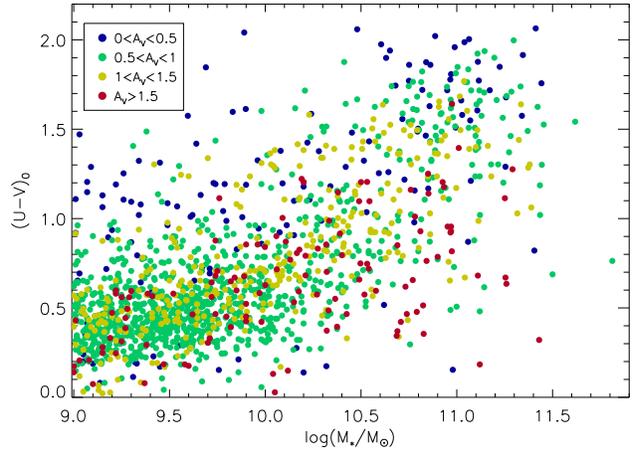}
\end{center}
\caption[]{Extinction corrected rest-frame U-V colour as a function of the stellar mass, with colour coding for A$_V$.\label{UVcorr-mass}}
\end{figure}

Figure \ref{UVcorr-mass} shows the de-reddened rest-frame U-V colour, (U-V)$_0$, as a function of M$_*$ for the sources in the sample. 
The less massive galaxies (M$_*$ $\lesssim$ 10$^{10}$ M$_\odot$) concentrate in a narrow blue cloud around (U-V)$_0$ $\sim$ 0.4, while more massive galaxies show increasingly red (U-V)$_0$, albeit with a high dispersion.

The position of the blue cloud is consistent with that found by \citet{Cardamone10} for the extinction corrected U-V colour of the Extended Chandra Deep Field South galaxies at 0.8$<$$z$$<$1.2. Nevertheless, their red galaxies clump in a disconnected red sequence, while we find a smooth transition from blue to red, with a very significant number of sources having intermediate colours.

Finding the cause for this discrepancy is not straightforward. 
The work by \citet{Cardamone10} uses stellar templates from \citet{Maraston05} and a \citet{Kroupa01} IMF, while we use the STELIB stellar library \citep{Bertelli94} assuming a \citet{Salpeter55} IMF. In principle, our A$_V$ estimates are sensitive to the selection of SPS models, which determine the shape and location of the DFS. However, changes in the DFS are negligible when switching to the models from \citet{Maraston05}, because differences between the two sets of models are important in the near-infrared, but not so much between the $U$ and $V$ bands.

The method used to estimate the amount of extinction may have a larger impact on the outcome. In the work of \citet{Cardamone10}, they use FAST \citep{Kriek09} to estimate the stellar mass, star-formation time scale, star-formation rate, and A$_V$ by fitting the observed SED with single-burst SPS models. Although the near-infrared photometry helps to break the age-extinction degeneracy, their A$_V$ estimate is influenced by all the stars in the galaxy, and not just those that dominate the emission in the $U$ and $V$ bands. If the extinction affecting different stellar populations within the galaxy varies with their age or metallicity, it is conceivable that the de-reddened U-V colours could be biased.

We also find that galaxies with M$_*$ $\gtrsim$ 10$^{10}$ M$_\odot$ show a correlation between (U-V)$_0$ and A$_V$: the more obscured sources also have bluer de-reddened U-V colours. This is consistent with the expectation of younger starforming galaxies being more dusty than quiescent galaxies, but other mechanisms could be responsible for this trend. For instance, if the amount of extinction is somehow overestimated then we will over-correct the U-V colour, obtaining the same correlation. 

An independent confirmation of the validity of A$_V$ estimates with the method of projection into the DFS can be obtained using the rest-frame V-J colour.
This colour has a smaller dependency on the age of stellar populations compared to U-V, and a larger one on extinction (since the V band is much more affected by extinction compared to the J band).
Because of this, it has been widely used in combination with another restframe colour, in particular U-V, to break the age-extinction degeneracy \citep[e.g.][]{Wuyts07,Williams09,Cardamone10,Brammer11}.

\begin{figure}
\begin{center}
\includegraphics[width=8.5cm]{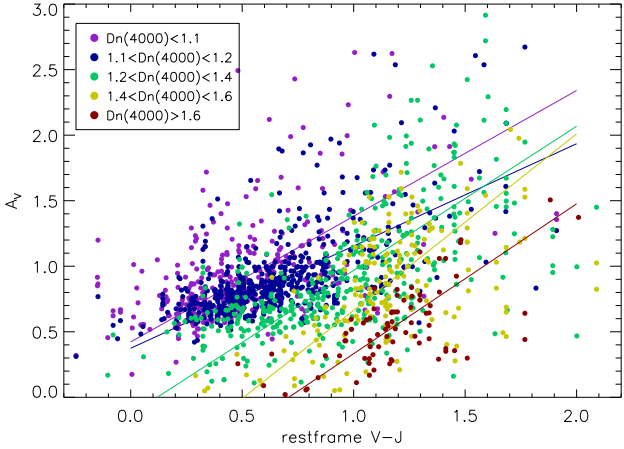}
\end{center}
\caption[]{Attenuation in the rest-frame V band as a function of the rest-frame V-J colour. Colours identify subsamples according to their \Dnu values. The continuous lines represent the bisector of the best-fitting linear relations for A$_V$ vs V-J and V-J vs A$_V$ for each subsample.\label{Av-VJ}}
\end{figure}

In Figure \ref{Av-VJ} we show A$_V$ versus rest-frame V-J, with separate colours identifying sources in distinct intervals of \Dnu.
The correlation between A$_V$ and V-J is evident only when considering galaxies with comparable \Dn values, because the dependency of the V-J colour with age is small but not negligible, and the typically lower extinction affecting older stellar populations compensates to some extent their redder intrinsic V-J colour.
Given the large dispersion, we take the bisector of the linear fits of A$_V$ vs V-J and V-J vs A$_V$ as the best linear model for the relationship between the two variables (solid lines in Figure \ref{Av-VJ}). 

Assuming a \citet{Draine03} extinction law with Milky Way grain size distribution \citep{Weingartner01} and R$_V$ = 3.1, the attenuation in the rest-frame J band is A$_J$ $\sim$ 0.3A$_V$. Therefore, the theoretical expectation for the slope of the best fitting linear model is $\frac{A_V}{E(V-J)}$ $\sim$ 1.4. The observed values range from 0.75 to 1.31 with a mean of 1.05, indicating that in spite of their large uncertainties, our A$_V$ estimates are broadly consistent with the observed restframe V-J colour.

\subsection{Average stellar ages}

Real galaxies have complicated SFHs, and finding a single value representing a characteristic stellar age for the entire stellar population is not straightforward. Furthermore, the dependency of \Dn and (U-V)$_r$ with metallicity for old ($>$1Gyr) stellar populations implies that such age estimates are metallicity-dependent (at least for old populations).

For convenience, we define the light-weighted average stellar age of a galaxy, $t_{ssp}$,  as the age of a solar-metallicity SSP that has a \Dn equal to the extinction corrected \Dn of the galaxy.
While the less massive galaxies in the sample are expected to have sub-solar metallicity, this does not invalidate $t_{ssp}$ estimates since (U-V)$_r$ and \Dn are nearly insensitive to metallicity for the typical ages of low mass galaxies (see Figure \ref{models}). On the other hand, for galaxies with old stellar populations and sub-(super-)solar metallicity, $t_{ssp}$ will under-(over-)estimate the actual age of their stellar populations.

Values of $t_{ssp}$ may differ significantly from the mean or the median age of individual stars in the galaxy. In particular, a recent burst of star formation can easily outshine the older stellar population in the UV-optical range and make the entire galaxy appear young, even if young stars only represent a small fraction of the stellar mass in the galaxy. Therefore, we interpret $t_{ssp}$ only as an indicator of the recent SFH of the galaxies.

\section{Discussion}

\subsection{Comparison to the local Universe}

\begin{figure}
\begin{center}
\includegraphics[width=8.5cm]{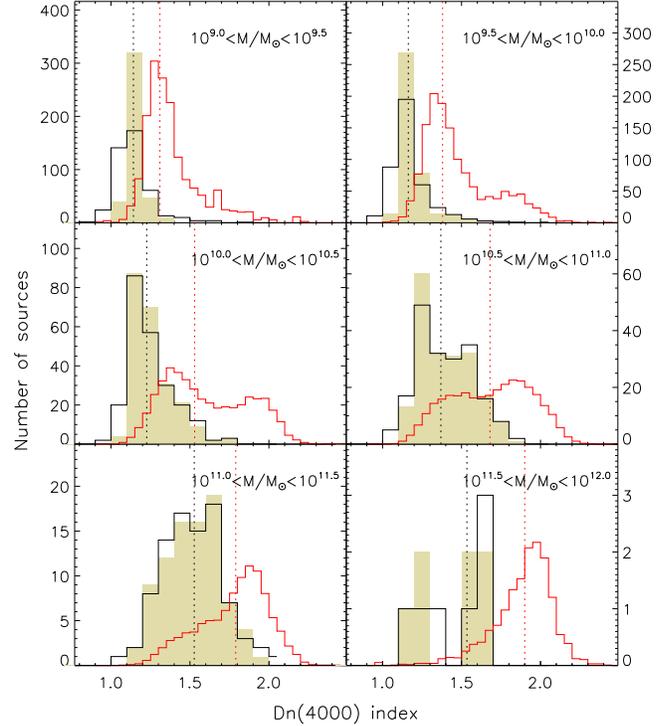}
\end{center}
\caption[]{Distribution of \Dn values for several stellar mass intervals. The black line represents the observed \Dn values, while the solid histogram represents their maximum likelihood estimates based on the method described in \S\ref{ssec-extinction-correction}. For comparison, the distribution found by \citet{Kauffmann03a} in SDSS galaxies is also shown (red histogram). Dotted lines indicate the median value of each distribution.\label{compare-Kauffmann}}
\end{figure}

In a large flux-limited sample of local galaxies from the SDSS, \citet{Kauffmann03b} found a sharp transition in the physical properties of galaxies at M$_*$ = 10$^{10.5}$ M$_\odot$, with galaxies below that value having lower surface mass densities, lower concentration indices, and younger stellar populations compared to galaxies above it.

Figure \ref{compare-Kauffmann} shows the distribution of \Dn values in bins of mass 0.5 dex wide for both their local sample and our intermediate redshift one. The \Dn distributions for the 0.65$<$z$<$1.07 galaxies peak at lower values compared to the local sample in all the mass intervals. Median values are also consistently lower, indicating significant evolution between these redshifts and the present epoch.

\begin{figure}
\begin{center}
\includegraphics[width=8.5cm]{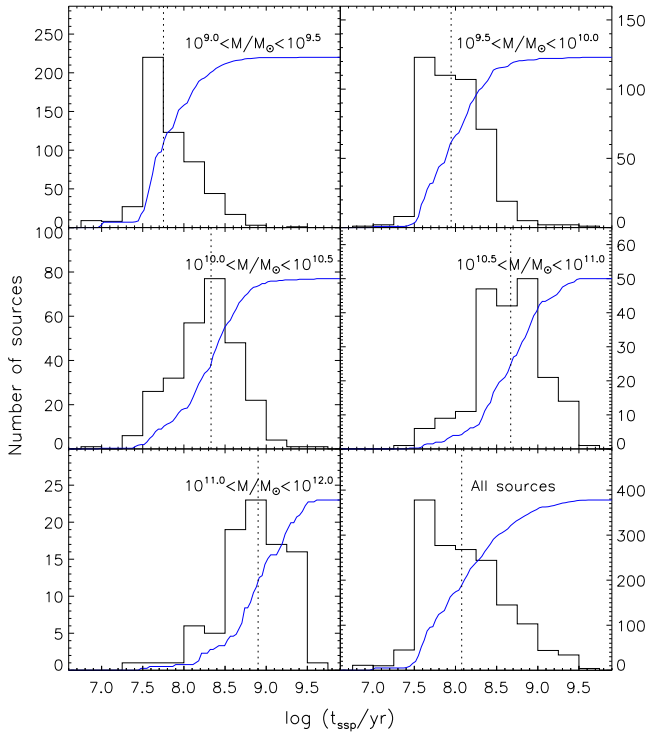}
\end{center}
\caption[]{Distribution of light-weighted stellar ages for several stellar mass intervals. The histogram shows the number of sources in intervals of $\log t_{ssp}$ of width 0.25, while the continuous curve shows the cumulative fraction of sources younger than a given age. The dotted line marks the median (50\% cumulative fraction) of each distribution. \label{tssp-histogs}}
\end{figure}

The bimodal distribution of \Dn values observed for local galaxies is not well developed in our intermediate redshift sample, and we can only marginally resolve the two peaks in the 10$^{10.5}$$<$M$_*$/M$_\odot$$<$10$^{11}$ mass bin (see Figures \ref{compare-Kauffmann} and \ref{tssp-histogs}). This is because even the most massive galaxies in our sample have not had enough time to evolve sufficiently their stellar populations.
Furthermore, while in the local sample there is a significant population of quiescent galaxies for all mass bins except the lowest one, in the 0.65$<$z$<$1.07 sample we find high \Dn values only in galaxies above 10$^{10.5}$ M$_\odot$, and a significant population of relatively young galaxies even in the highest mass ranges.

In the local sample, the relative abundances of young and old galaxies vary slowly with the stellar mass, while in the 0.65$<$z$<$1.07 sample we find a relatively abrupt  transition at M$_*$$\sim$10$^{10.5}$ M$_\odot$ from a distribution dominated by young populations to another where there are comparable numbers of galaxies with young and old stars.

\subsection{The mass-dependency of extinction}\label{ssec-Av-mass}

Studies with large spectroscopic samples from the SDSS have shown that the fundamental property governing extinction of the H$_\alpha$ emission in star-forming galaxies is the stellar mass \citep{Garn10,Zahid12,Zahid13}. 
Here we discuss whether the stellar mass also influences the extinction affecting the stellar population.

\begin{figure}
\begin{center}
\includegraphics[width=8.5cm]{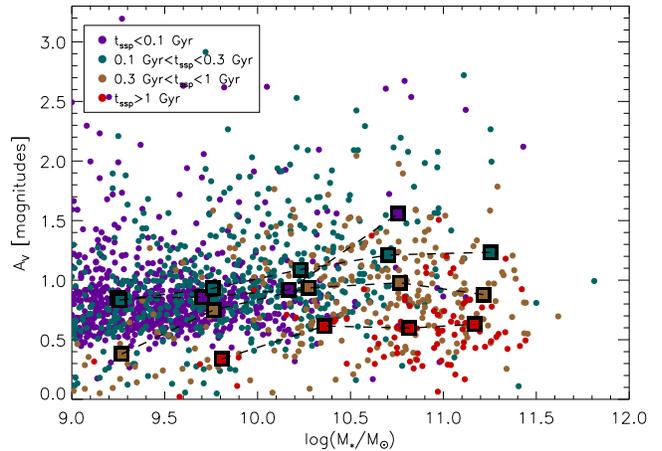}
\end{center}
\caption[]{Apparent absorption in the rest-frame V band versus stellar mass, with colour coding for the light-weighted stellar age. Big squares represent mean values of M$_*$ and A$_V$ in bins of stellar mass 0.5 dex wide.\label{Av-mass}}
\end{figure}

Figure \ref{Av-mass} shows A$_V$ as a function of M$_*$ in four intervals of $t_{ssp}$. Most galaxies concentrate in a relatively narrow band of A$_V$ values between 0.5 and 1.0,
and there is not any significant trend with mass for the sample as a whole other than an increase in dispersion in the 10$^{10}$--10$^{11}$ M$_\odot$ range.
Nevertheless, when sources are grouped by their light weighted stellar ages, and mean A$_V$ values are calculated in bins of stellar mass, a pattern emerges in which the mean A$_V$ for a given age interval increases with the stellar mass, reaching a maximum in the 10--10.5 or 10.5--11 mass bin, and stabilises or decreases at higher masses. 

The sources with A$_V$$>$1.5 concentrate in the mass range 10$^{10}$--10$^{11}$ M$_\odot$, with moderate \Dn values indicating relatively young stellar populations and red V-J confirming high extinction. These galaxies are among the most massive of the (intrinsic) blue galaxies (see Figure \ref{UVcorr-mass}), and their mass distribution is strongly biased against lower mass galaxies compared to less obscured sources.

For a given mass interval, the mean A$_V$ decreases with increasing stellar age. This trend is stronger among the massive galaxies (M$_*$$>$10$^{10.5}$ M$_\odot$), where the average A$_V$ is $>$1.2 magnitudes for $t_{ssp}$$<$0.3 Gyr but only $\sim$0.6 magnitudes for $t_{ssp}$$>$1 Gyr. While this is consistent with expectations of lower extinction in quiescent galaxies, we caution that an older $t_{ssp}$ is also the expected result if A$_V$ is somehow underestimated.

\subsection{Constraining star formation histories}

The light-weighted average stellar age does not provide information about the entire SFH of the galaxies. In particular, for galaxies experiencing a recent burst of star formation, the pre-burst SFH cannot be inferred from these data, since emission from hot young stars outshines the old population and produces (U-V)$_r$ and \Dn values very close to those of a SSP with the age of the burst. However, we can still set some interesting constraints on the SFH of most galaxies.
The track for constant SFR in Figure \ref{Dn4000-UV-tracks} overlaps with that of a SSP 0.3 Gyr old. This implies that galaxies with $\log t_{ssp}$$<$8.5 have experienced a recent ($<$300 Myr) increase in their SFR, while those with larger values are experiencing a decline in their SFR at least in the last few hundred Myr.

\begin{figure}
\begin{center}
\includegraphics[width=8.5cm]{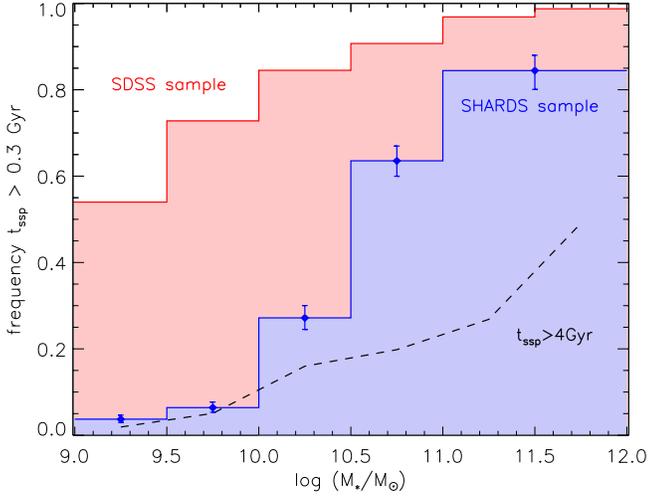}
\end{center}
\caption[]{Fraction of sources with declining SFR ($t_{ssp}>$0.3 Gyr) as a function of stellar mass, for the local SDSS sample (red histogram) and the SHARDS sample (blue histogram). Error bars indicate the 68\% confidence interval calculated using the Wilson formula for binomial distributions.  The dashed line shows the fraction of SDSS sources with $t_{ssp}>$4Gyr, compatible with passive evolution since at least $z$$\sim$0.65.\label{red-sequence}}
\end{figure}

Figure \ref{red-sequence} shows the fraction of galaxies with $t_{ssp}>$0.3 Gyr (that is, with declining SFR) as a function of M$_*$ for both our sample and the local SDSS sample of \citet{Kauffmann03a}. In the latter we assumed solar metallicity and a typical extinction of A$_V$=1 to convert their \Dn values to $t_{ssp}$ ($t_{ssp}$=0.3 Gyr corresponds to \Dnu=1.3). 

The fraction of sources with $t_{ssp}>$0.3 Gyr increases steadily with the stellar mass in the SHARDS sample, from $\sim$5\% in the lowest mass bin to $\sim$80\% in the highest.
This indicates that at $z$$\sim$0.9 star formation is clearly in decline or halted in the most massive galaxies, but it is still increasing for galaxies below 10$^{10.5}$ M$_\odot$, in qualitative agreement with the downsizing paradigm.
On the other hand, in the local SDSS sample the SFR is predominantly in decline except perhaps in the lowest mass bin, and the fraction of $t_{ssp}>$0.3 Gyr sources reaches levels close to 100\% in the most massive bin. 

Note, however, that a decline in SFR does not necessarily imply that these galaxies will undergo passive evolution down to $z$$\sim$0. If we were to start from the 0.65$<$$z$$<$1.07 sample and apply passive evolution during the next 5--7 Gyr, their \Dn distribution at $z$$\sim$0.1 (the median redshift of the SDSS sample) would be pushed to the right and compressed in the interval \Dnu$\sim$1.9--2.1. Clearly, this is not what we see in the SDSS sample (Figure \ref{compare-Kauffmann}), albeit one half of the galaxies in the highest mass bin have \Dnu$>$1.9, equivalent to $t_{ssp}$$>$4 Gyr.

Interestingly, the fraction of $t_{ssp}$$>$0.3 Gyr sources in the SHARDS sample is roughly twice the fraction of $t_{ssp}$$>$4 Gyr sources in the SDSS sample in all the mass bins. When accounting for the doubling in the comoving density of galaxies with 10$^{10}$$<$ M$_*$/M$_\odot$$<$10$^{11.5}$ since $z$$\sim$0.9 [PG08], we have roughly one quiescent galaxy with $t_{ssp}$$>$4 Gyr in the SDSS sample for every galaxy with declining SFR in the SHARDS sample. This suggests that most of these intermediate-$z$ galaxies might never reactivate their star formation but instead evolve passively until the present time.

In the intermediate mass bins (10$^{10}$$<$M$_*$/M$_\odot$$<$10$^{11}$), about 50\% is increasing its SFR and the other 50\% decreasing in the SHARDS sample. Their distribution peaks close to $\log t_{ssp}$=8.5, which may be interpreted in two ways: either most of the galaxies in this mass range sustain relatively stable SFRs, or they undergo repeated episodes of star formation, separated by more quiescent periods no longer than a few hundred Myr. 

The highest \Dn values observed in the sample, $\sim$1.8, correspond to $t_{ssp}$$\sim$3-4 Gyr or $\tau$=1 Gyr for a 6 Gyr old galaxy with an exponentially declining SFR. In any case, this implies that more than 80\% of the stellar mass in these galaxies was already in place 4 Gyr earlier (that is, at $z$$\sim$2--4) than the epoch observed here.
However, these very old galaxies represent only a very small fraction of the massive galaxies in the sample. When considering all the galaxies with M$_*$$>$10$^{11}$ M$_\odot$, only 10\% of them has $t_{ssp}$$>$2.5 Gyr.

\section{Conclusions}

We have presented the analysis of the stellar populations of a mass-selected sample of galaxies at $z$$\sim$0.9 with ultra-deep (m$_{AB}$$<$26.5) optical medium-band (R$\sim$50) photometry 
from the Survey for High-z Absorption Red and Dead Sources (SHARDS).
We demonstrate that the spectral resolution of SHARDS allows for a consistent measure of the \Dn index for all galaxies at 0.65$<$$z$$<$1.07 down to M$_*$=10$^9$ M$_\odot$, roughly 1/10$^{th}$ the threshold of similar studies based on spectroscopy.

The stacked SHARDS SEDs of sources grouped by their stellar mass show increasingly red continua, stronger 4000 \AA{} breaks, and weaker \oii emission with increasing stellar mass, suggesting that stellar age, and not extinction, is the dominant factor driving the correlation between optical colours and stellar mass.
When considering individual sources, both \Dn and (U-V)$_r$ correlate with M$_*$. The dispersion in \Dn values at a given M$_*$ increases with M$_*$, while for (U-V)$_r$ decreases due to the higher average extinction prevalent in massive star-forming galaxies.

We find a small but significant evolution with redshift of the restframe U-V colour within the sample. Galaxies at $z$$\sim$0.7 are between 0.1 and 0.2 magnitudes redder compared to galaxies at $z$$\sim$1.0 in the same mass range. The stronger evolution occurs at intermediate masses (10$^{9.5}$--10$^{10.5}$ M$_\odot$).

We present a new method for obtaining A$_V$ estimates based on the differences in sensitivity of the (U-V)$_r$ and $\log$\Dn colours to extinction. This method allows us to break the degeneracy between age and extinction, and to produce extinction-corrected observables.

The extinction corrected U-V colour of blue cloud galaxies in our sample is consistent with other studies at similar redshifts. Nevertheless, for massive galaxies we find a smooth transition towards the red sequence, with many sources at intermediate colours, in contrast to the strong bimodality found by \citet{Cardamone10}. We interpret the discrepancy as due to differences in the methods used to estimate the extinction affecting the stellar population.

Compared to local galaxies, the distributions of \Dn for our sample peak at lower values indicative of younger stellar populations for all mass ranges. The bimodal distribution of \Dn is rather absent at $z$$\sim$0.9 contrarily to the stark one observed in local galaxies, because even the most massive, already quiescent galaxies have not had enough time to evolve sufficiently their stellar populations. We find a relatively sharp transition at M$_*$$\sim$10$^{10.5}$M$_\odot$ from a distribution dominated by young stellar populations to another where there are comparable numbers of galaxies with young and old stars.

When galaxies are grouped by their light weighted stellar ages, we find a weak positive correlation between extinction and stellar mass, which probably arises from the well-known correlation between stellar mass and metallicity.

The fraction of SHARDS sources where star formation is in decline increases steadily with the stellar mass, from $\sim$5\% at $\log$(M$_*$/M$_\odot$)=9.0--9.5 to $\sim$80\% at $\log$(M$_*$/M$_\odot$)$>$11. This indicates that at $z$$\sim$0.9 star formation is clearly in decline or halted in the most massive galaxies, but it is still increasing for galaxies below 10$^{10.5}$, in qualitative agreement with the downsizing paradigm. 

For the more massive galaxies, the comoving density of $z$$\sim$0.9 galaxies with declining SFR is comparable to the density of local galaxies with old ($t_{ssp}$$>$4Gyr) stellar populations, suggesting that most of these galaxies never re-activated their star formation. 

The higher \Dn values observed in the SHARDS sample correspond to $t_{ssp}$$\sim$3--4Gyr, and imply that at least 80\% of the stellar mass in these galaxies was already in place at $z$$\sim$2--4. However, these very old galaxies represent only a small fraction ($<$10\%) of the M$_*$$>$10$^{11}$ M$_\odot$ galaxies.

\section*{Acknowledgements}
A.H.-C. and A.A.-H. acknowledge funding by the Universidad de Cantabria Augusto Gonz\'alez Linares program.
We acknowledge support from the Spanish Programa Nacional de
Astronom\'{\i}a y Astrof\'{\i}sica under grants AYA2009-07723-E and AYA2009-10368. SHARDS has been funded by the Spanish MICINN/MINECO under the Consolider-Ingenio 2010 Program grant CSD2006-00070: First Science with the GTC. This work has
made use of the Rainbow Cosmological Surveys Database, which is operated by the Universidad Complutense de Madrid (UCM).
Based on observations made with the Gran Telescopio Canarias (GTC), installed at the Spanish Observatorio del Roque de los
Muchachos of the Instituto de Astrof\'{\i}sica de Canarias, in the island of La Palma. We thank all the GTC Staff for their support
and enthusiasm with the SHARDS project, and we would like to especially acknowledge the help from Antonio Cabrera and Ren\'e Rutten. We thank the anonymous referee for their useful comments that helped to improve this paper.

\onecolumn

\begin{deluxetable}{c r r r r r r r} 
\tabletypesize{\scriptsize}
\tablecaption{General sample properties\label{general-table}}
\tablewidth{0pt}
\tablehead{\colhead{Mass range} & \colhead{N total} & \colhead{N $z_{\rm{spec}}$} & \colhead{N \Dnu} &
 \colhead{$\langle\log$(M$_*$/M$_\odot$)$\rangle$} & \colhead{$\langle z \rangle$} & \colhead{$\langle$(U-V)$_r$$\rangle$} & 
 \colhead{$\langle$\Dnu$\rangle$}}
\startdata
 9.0--9.5 & 592 & 163 & 549 &  9.25 &  0.87 &  0.96 &  1.14 \\
 9.5--10.0 & 475 & 243 & 457 &  9.73 &  0.89 &  1.10 &  1.18 \\
10.0--10.5 & 279 & 192 & 277 & 10.23 &  0.89 &  1.40 &  1.26 \\
10.5--11.0 & 204 & 160 & 204 & 10.76 &  0.91 &  1.79 &  1.42 \\
11.0--12.0 &  94 &  79 &  91 & 11.20 &  0.90 &  1.90 &  1.52 \\
Total  &1644 & 837 &1578 &  9.85 &  0.89 &  1.23 &  1.23 \\

\enddata

\end{deluxetable}

\twocolumn

\appendix
\section{Calibration of \Dn measurements using zCOSMOS spectra}
\subsection{Simulation of SHARDS photometry}

We simulated SHARDS photometry by convolving a library of optical spectra with the transmission profiles of the SHARDS filters. The library was selected from the zCOSMOS \citep{Lilly07} Data Release 2, and includes the 1377 sources in the 0.5 $<$ $z$ $<$ 1.2 range with reliable redshift determinations that are not classified as AGN and are brighter than magnitude 22.5 (AB) in the $i$ band. The cut in optical magnitude ensures the spectrum has a well detected continuum and the 4000 \AA{} break can be properly measured in the full resolution spectrum.

Photometric errors are simulated by adding to each magnitude an error term m$_{\rm{err}}$ = 0.1$\sigma$, where $\sigma$ is a random variable with normal distribution. This roughly corresponds to the 1-$\sigma$ uncertainty of the SHARDS photometry at AB magnitude $\sim$25.5, and is higher than the actual photometric errors for $\sim$85--90\% of the SHARDS sources. 

\begin{figure}
\begin{center}
\includegraphics[width=8.5cm]{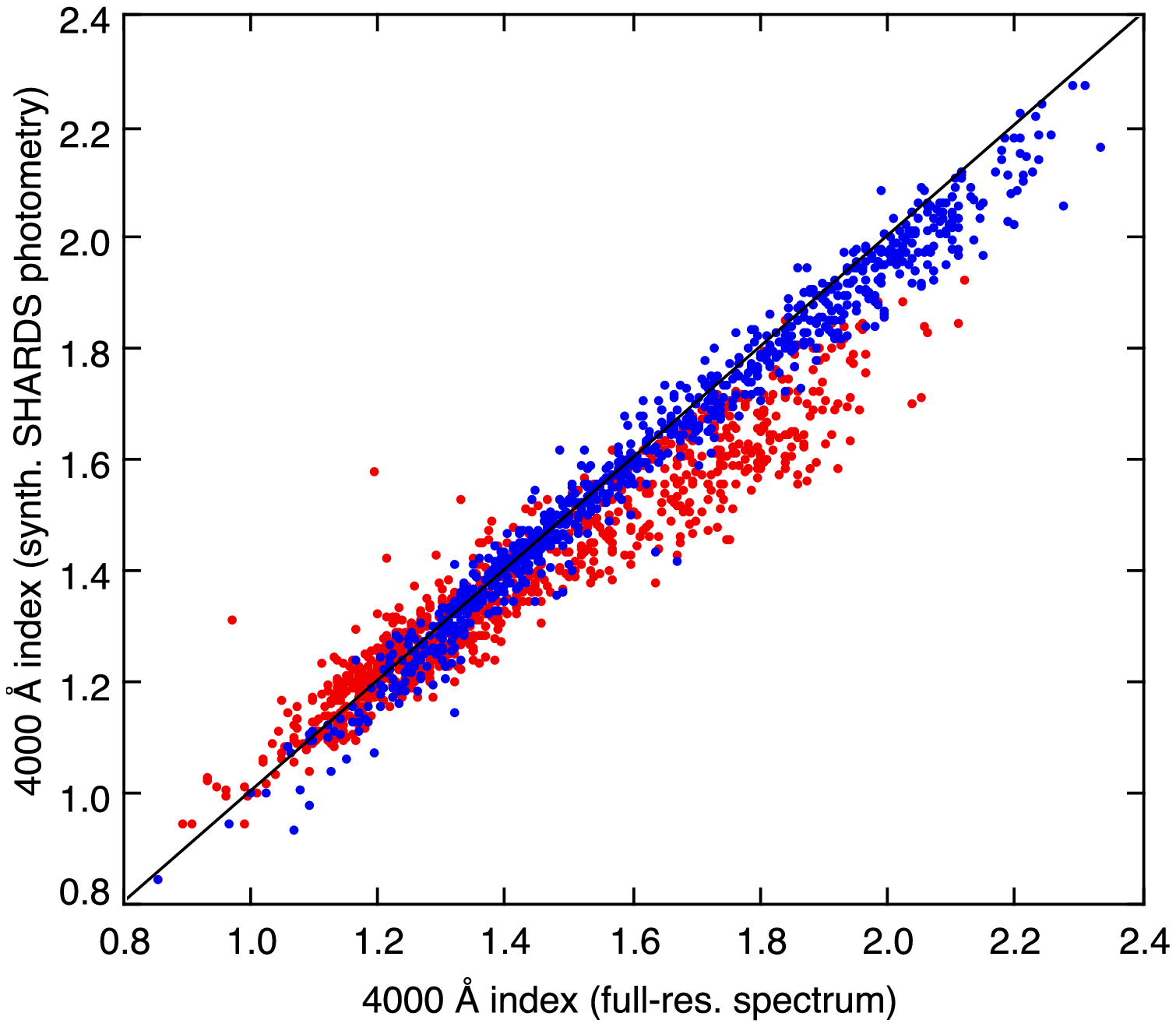}
\end{center}
\caption[]{Comparison between 4000 \AA{} indices for simulated SHARDS photometry and those from full resolution zCOSMOS spectra. The blue dots represent measurements of the ``regular'' D(4000) index, while the red ones represent the ``narrow'' version \Dnu. \label{D4000-zCOSMOS}}
\end{figure}

The D(4000) and \Dn indices are measured in the simulated photometry using the same procedure employed for the SHARDS data. 
Figure \ref{D4000-zCOSMOS} compares values of the D(4000) and \Dn indices measured directly on the full resolution zCOSMOS spectra (D4$_C$ and Dn4$_C$) with those recovered from the synthetic photometry (D4$_P$ and Dn4$_P$). 

The dispersion in D4$_P$ is markedly uniform in the entire range of 4000 \AA{} break strengths, at about 5\% of its average value, while for Dn4$_P$ the dispersion is significantly higher, and also has stronger bias towards lower values of the index in the simulated photometry compared to the full resolution spectrum. 

\subsection{Corrected \Dn values}

Most recent studies prefer the narrow definition of the 4000 \AA{} index due to its lower dependency on extinction. This is thus the most convenient definition for comparison with results from other samples. Nevertheless, we just showed that at the spectral resolution of the SHARDS SED (R$\sim$50) the measurement of \Dn is significantly more affected by the interpolation error when compared to D(4000).

To overcome this issue, we calculate \Dn values corrected for the interpolation bias from the D(4000) measurements on the SHARDS photometry. The correction terms are obtained from polynomial fitting of Dn4$_C$ values as a function of D4$_P$ in the zCOSMOS sample.
To this aim, the individual zCOSMOS sources are sorted by their D4$_P$ values, and the mean and standard deviation of their D4$_P$ and Dn4$_C$ values and are calculated in bins of 30 sources each. 

\begin{figure}
\begin{center}
\includegraphics[width=8.5cm]{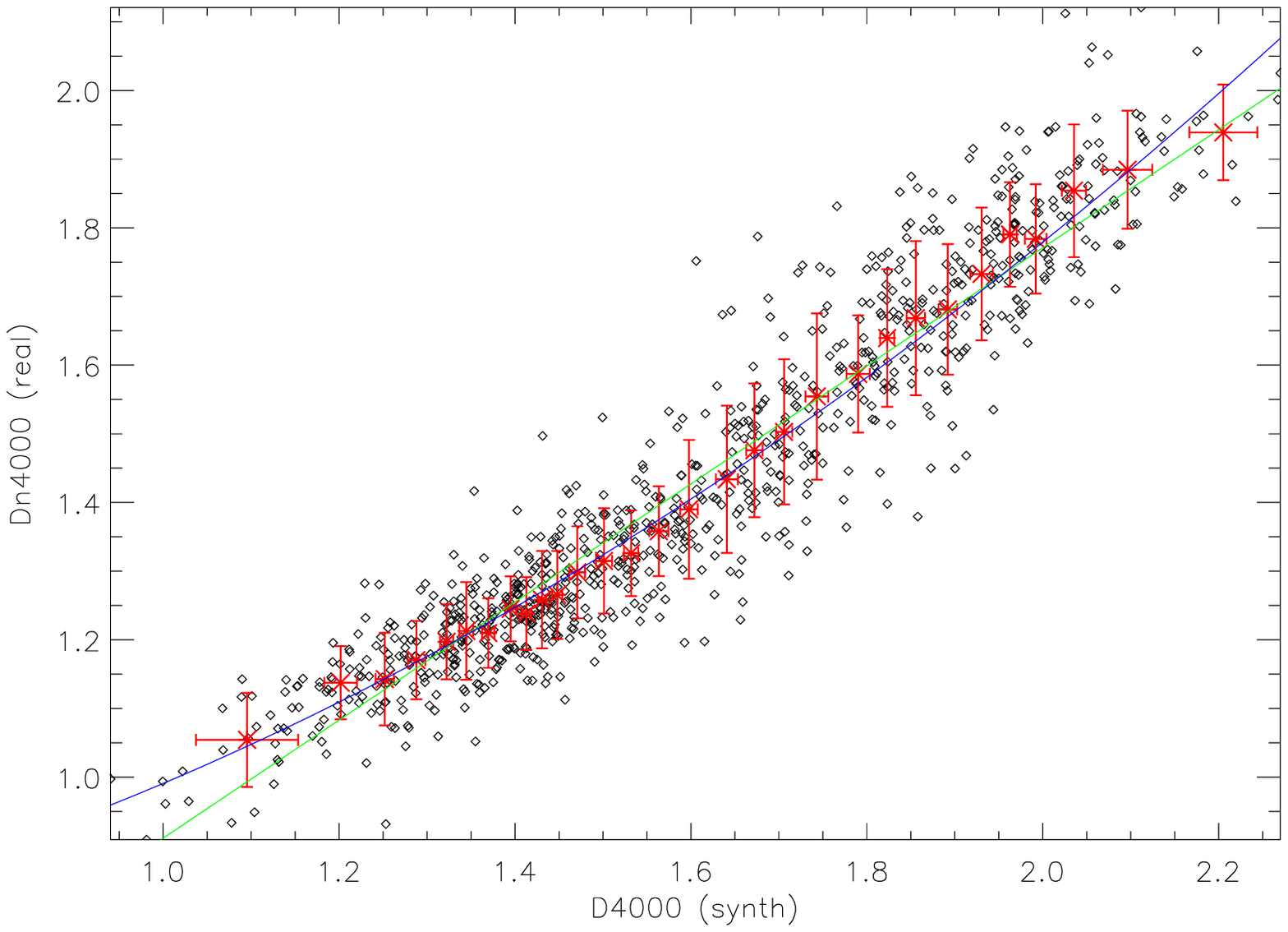}
\end{center}
\caption[]{Comparison between \Dn values measured in the full resolution spectrum and D(4000) values measured in the synthetic SHARDS photometry. Black dots represent individual zCOSMOS sources, while red asterisks mark the average values in bins of 30. Error bars indicate the 1-$\sigma$ dispersion in each bin. The green and blue lines represent the best fitting linear and parabolic models, respectively.\label{D4000-regresion-fit}}
\end{figure}

Figure \ref{D4000-regresion-fit} shows Dn4$_C$ versus D4$_P$ and the best fitting linear and parabolic models. In both cases, the expected values, Dn4$_M$, are comfortably within one standard deviation from the observed ones, Dn4$_C$, with the only exception of the lowest and highest D4$_P$ bins for the linear and parabolic models, respectively. 
Nevertheless, the linear model fails to predict the convergence of D4$_P$ and Dn4$_C$ values at D4$_P$ = Dn4$_C$ = 1, which is caused by the index measurement becoming insensitive to the spectral resolution in sources with a flat spectrum near 4000 \AA. 
On the other hand, the parabolic model reproduces this feature accurately.

\begin{figure}
\begin{center}
\includegraphics[width=8.5cm]{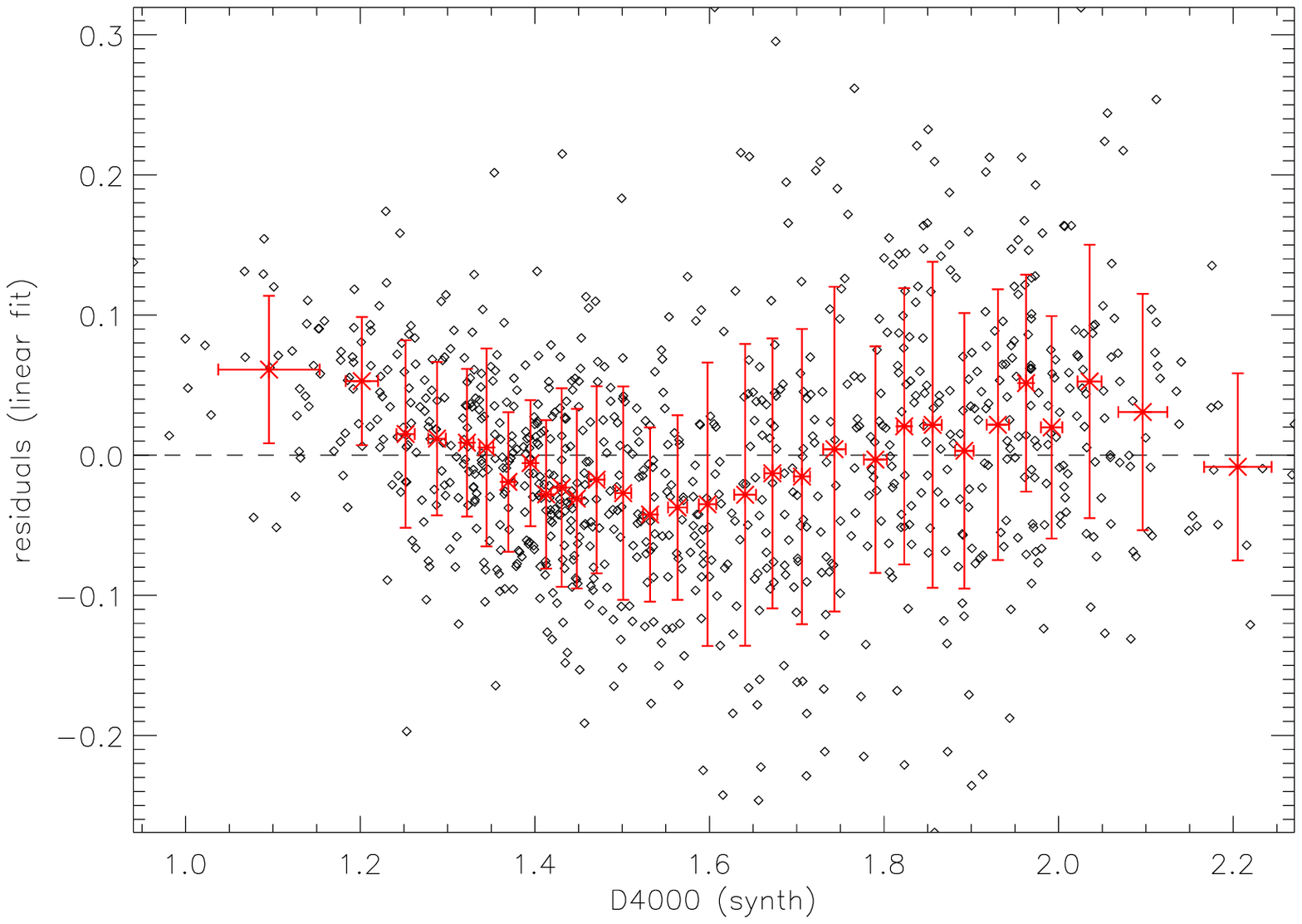}
\includegraphics[width=8.5cm]{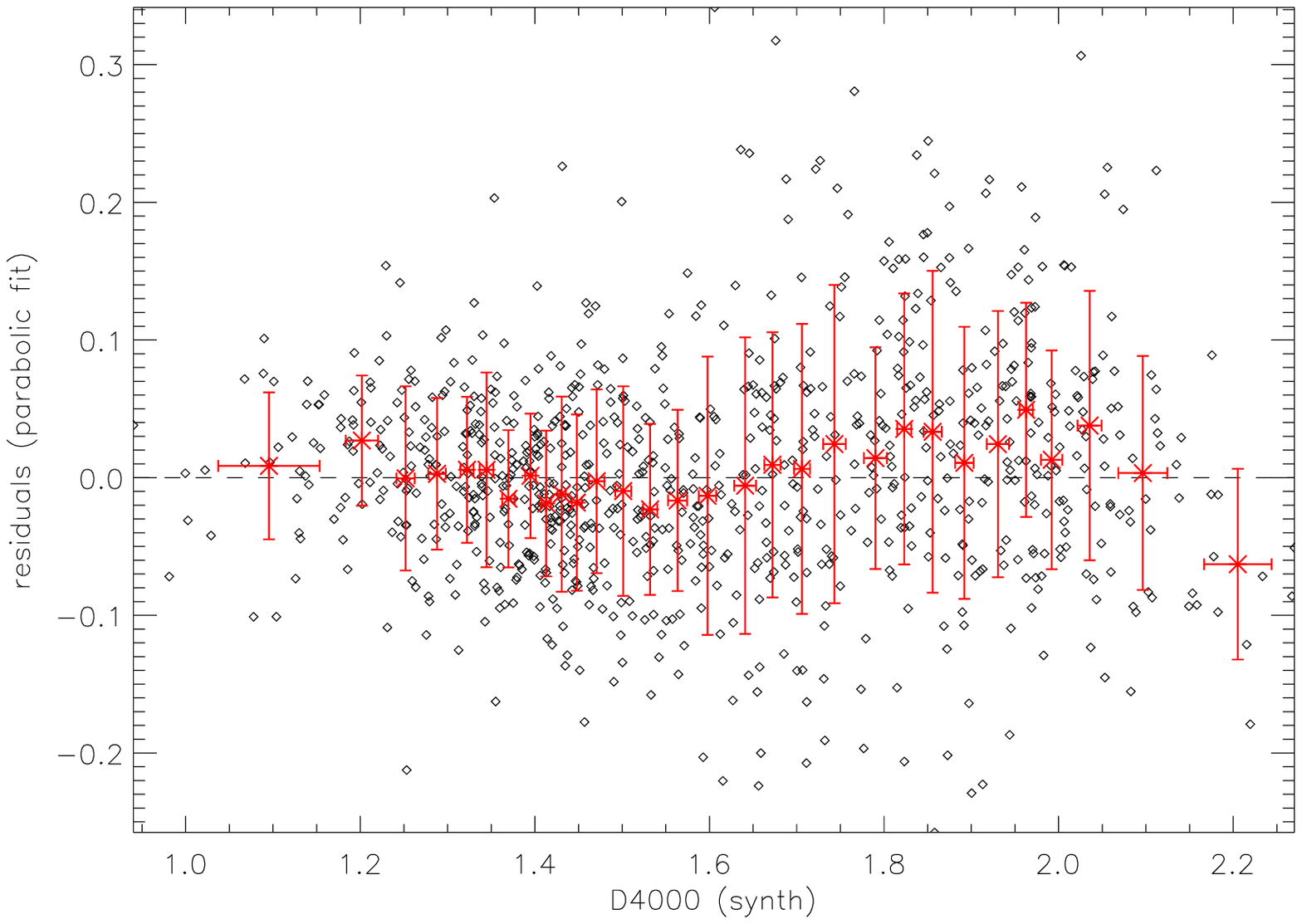}
\end{center}
\caption[]{Residuals in \Dn after subtraction of the best fitting linear (top) and parabolic (bottom) models. Symbols as in Figure \ref{D4000-regresion-fit}.\label{D4000-residuals-fit}}
\end{figure}

The residuals ($\delta_M$ = Dn4$_C$ - Dn4$_M$) show that the parabolic fit reproduces much better the observed values except in the 1.8$<$D4$_P$$<$2.1 range, where both models predict slightly lower values than observed (Figure \ref{D4000-residuals-fit}). Because of this issues, the parabolic model is preferred over the linear one.

From the coefficients of the parabolic model we obtain the correction term, $\Delta$, that needs to be subtracted from the D(4000) values measured in the SHARDS photometry to obtain corrected \Dn values:

\begin{equation}
\Delta = 0.482(D4_P - 1) - 0.263(D4_P - 1)^2
\end{equation}
which implies corrections of 0.00, 0.17, and 0.22 at D4$_P$ = 1.0, 1.5, and 2.0, respectively.

The systematic error in Dn4$_M$ introduced by the parabolic model is at most $\sim$3\%, and the 1-$\sigma$ dispersion of Dn4$_M$ values for a given Dn4$_C$ is $\sim$5\%. 

\subsection{Dependency with the photometric errors}

\begin{figure}
\begin{center}
\includegraphics[width=8.5cm]{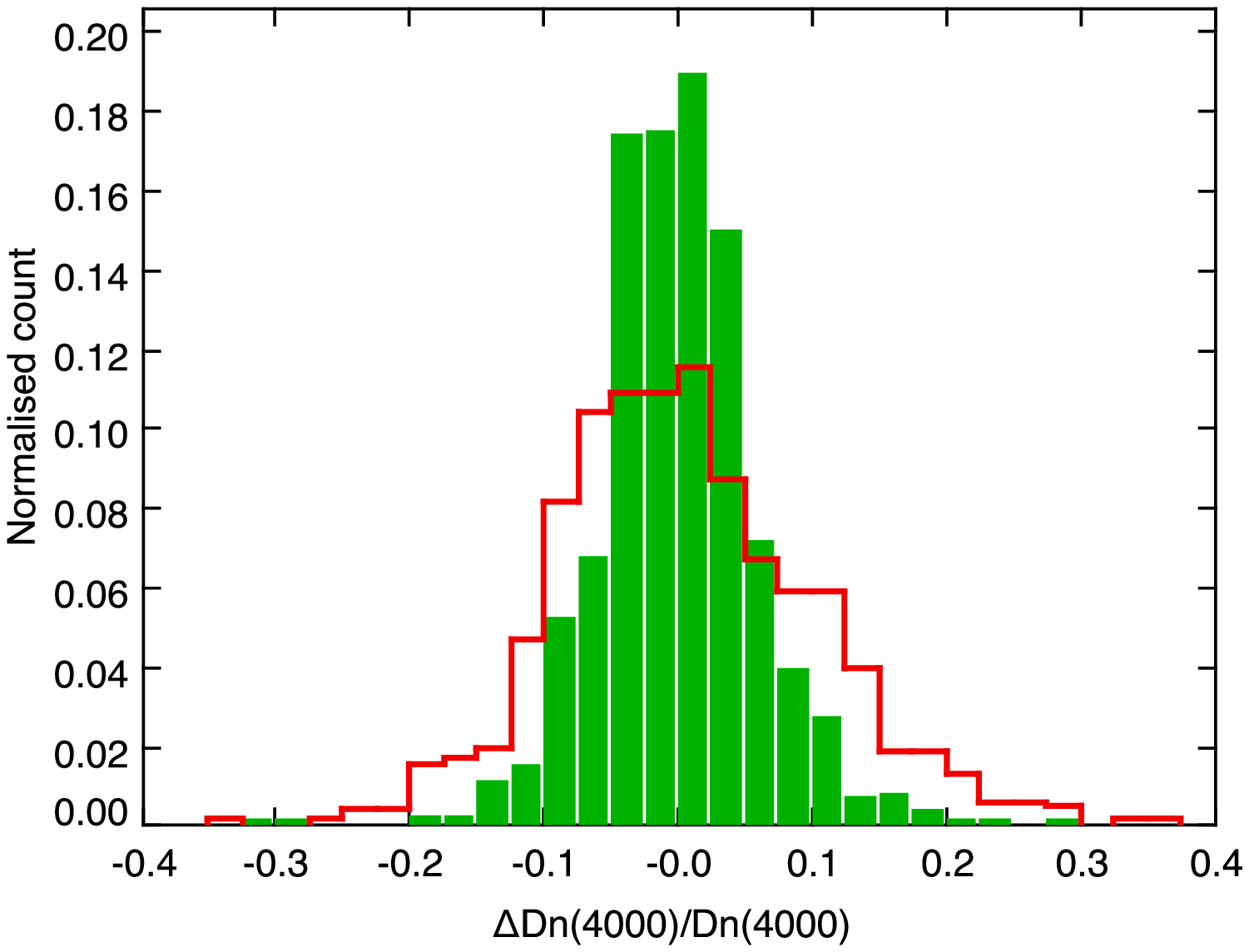}
\end{center}
\caption[]{Distribution of relative errors in the \Dn index measurement for simulated SHARDS data with photometric errors $\Delta$m = 0.01 (green bars) and $\Delta$m = 0.1 (red line).\label{delta-Dn4000-synth}}
\end{figure}

The calibration used to obtain corrected \Dn values assumes photometric uncertainties around 0.1 magnitudes, but in some sources (particularly the more massive galaxies) these can be an order of magnitude lower. 
Repeating the above calibration steps on simulated SHARDS data with 0.01 magnitudes uncertainties yields nearly identical calibration coefficients. Furthermore, the dispersion of residuals is only slightly lower (Figure \ref{delta-Dn4000-synth}), indicating that it is the interpolation error, and not the photometric errors, the main cause of dispersion in the Dn4$_M$ versus Dn4$_C$ relationship.

\end{document}